\documentclass[journal]{IEEEtran}

\ifCLASSINFOpdf
\else

\fi

\usepackage{amsmath,cite,multirow,amsfonts}
\usepackage{graphicx}

\usepackage{subfigure}
\newcommand\numberthis{\addtocounter{equation}{1}\tag{\theequation}}

\begin{document}

\title{Joint Optimization of Masks and Deep Recurrent Neural Networks for Monaural Source Separation}

\author{Po-Sen~Huang,~\IEEEmembership{Member,~IEEE,}
		Minje~Kim,~\IEEEmembership{Member,~IEEE,}
        Mark~Hasegawa-Johnson,~\IEEEmembership{Senior Member,~IEEE,}
        and~Paris~Smaragdis,~\IEEEmembership{Fellow,~IEEE}

\thanks{Manuscript received February 11, 2015; revised June 02, 2015; accepted July 11, 2015. Date of publication August 13, 2015; date of current version null-date. This work was supported by the U.S. Army Research Laboratory and Army Research Office under Grant W911NF-09-1-0383 and the National Science Foundation under Grant 1319708. This work used the Extreme Science and Engineering Discovery Environment (XSEDE), which is supported by the National Science Foundation under Grant ACI-1053575. The associate editor coordinating the review of this manuscript and approving it for publication was Prof. Mads Christensen.}
\thanks{P.-S.~Huang is with the Department of Electrical and Computer Engineering,~University of Illinois at Urbana-Champaign, Urbana,
	IL, 61801 USA, and also with Clarifai, New York, NY, 10010 (email: huang146@illinois.edu)}
\thanks{M.~Hasegawa-Johnson is with the Department
of Electrical and Computer Engineering,~University of Illinois at Urbana-Champaign, Urbana,
IL, 61801 USA (email: jhasegaw@illinois.edu)}
\thanks{M. Kim is with the Department of Computer Science, University of Illinois at Urbana-Champaign, Urbana, IL, 61801 USA (email: minje@illinois.edu)}
\thanks{P. Smaragdis is with the Department of Computer Science and Department of Electrical and Computer Engineering, University of Illinois at Urbana-Champaign, Urbana, IL, 61801 USA, and also with Adobe Research, San Francisco, CA, 94103, USA (email: paris@illinois.edu)}
\thanks{Digital Object Identifier 10.1109/TASLP.2015.2468583}
}

\markboth{IEEE/ACM TRANSACTIONS ON AUDIO, SPEECH, AND LANGUAGE PROCESSING, VOL. 23, NO. 12, DECEMBER 2015}%
{Huang \MakeLowercase{\textit{et al.}}: Joint Optimization of Ratio Mask with DRNN for Monaural Source Separation}


\maketitle

\begin{abstract} 
Monaural source separation is important for many real world applications. 
It is challenging because, with only a single channel of information available, without any constraints, an infinite number of solutions are possible. 
In this paper, we explore joint optimization of masking functions and deep recurrent neural networks for monaural source separation tasks, including monaural speech separation, monaural singing voice separation, and speech denoising.
The joint optimization of the deep recurrent neural networks with an extra masking layer enforces a reconstruction constraint. Moreover, we explore a discriminative criterion for training neural networks to further enhance the separation performance. 
We evaluate the proposed system on the TSP, MIR-1K, and TIMIT datasets for speech separation, singing voice separation, and speech denoising tasks, respectively. 
Our approaches achieve 2.30--4.98 dB SDR gain compared to NMF models in the speech separation task, 2.30--2.48 dB GNSDR gain and 4.32--5.42 dB GSIR gain compared to existing models in the singing voice separation task, and outperform NMF and DNN baselines in the speech denoising task.
\end{abstract}


 \ifCLASSOPTIONpeerreview
 \begin{center} \bfseries EDICS Category: 3-BBND \end{center}
 \fi
\IEEEpeerreviewmaketitle

\section{Introduction}\label{sec:introduction}
\IEEEPARstart{S}{ource} separation is a problem in which several signals have been mixed together and the objective is to recover the original signals from the combined signals. 
Source separation is important for several real-world applications. 
For example, the accuracy of chord recognition and pitch estimation can be improved by separating the singing voice from the music accompaniment \cite{Huang_RPCA_Separation_ICASSP2012}. The accuracy of automatic speech recognition (ASR) can be improved by separating speech signals from noise \cite{maas2012recurrent}. 
Monaural source separation, i.e., source separation from monaural recordings, is particularly challenging because, without prior knowledge, there is an infinite number of solutions. In this paper, we focus on source separation from monaural recordings with applications to speech separation, singing voice separation, and speech denoising tasks.

Several approaches have been proposed to address the monaural source separation problem. We categorize them into \textit{domain-specific} and \textit{domain-agnostic} approaches. 
For domain-specific approaches, models are designed according to the prior knowledge and assumptions of the tasks. For example, in singing voice separation tasks, several approaches have been proposed to exploit the assumption of the low rank and sparsity of the music and speech signals, respectively \cite{Huang_RPCA_Separation_ICASSP2012, Pablo_lowrank_2012, Yang_2013_LowReprOfBoth, Yang_sparse_lowrank_2012}. In speech denoising tasks, spectral subtraction \cite{Boll_spectralsub_1979} subtracts a short-term noise spectrum estimate to generate the spectrum of a clean speech. By assuming the underlying properties of speech and noise, statistical model-based methods infer speech spectral coefficients given noisy observations \cite{Ephraim_MMSE_1984}.
However, in real-world scenarios, these strong assumptions may not always hold. For example, in the singing voice separation task, the drum sounds may lie in sparse subspaces instead of being low rank. In speech denoising tasks, the models often fail to predict the acoustic environments due to the non-stationary nature of noise.


For domain-agnostic approaches, models are learned from data directly without having any prior assumption in the task domain. 
Non-negative matrix factorization (NMF) \cite{lee1999learning} and probabilistic latent semantic indexing (PLSI) \cite{hofmann1999probabilistic, smaragdis2006probabilistic} learn the non-negative reconstruction bases and weights of different sources and use them to factorize time-frequency spectral representations.
NMF and PLSI can be viewed as a linear transformation of the given mixture features (e.g. magnitude spectra) during the prediction time.
However, based on the minimum mean squared error (MMSE) estimate criterion, the optimal estimator $\mathbb{E} [ \mathbf{Y}|\mathbf{X} ]$ is a linear model in $\mathbf{X}$ only if $\mathbf{X}$ and $\mathbf{Y}$ are jointly Gaussian, where $\mathbf{X}$ and $\mathbf{Y}$ are the mixture and separated signals, respectively. 
In real-world scenarios, since signals might not always follow Gaussian distributions, linear models are not expressive enough to model the complicated relationship between separated and mixture signals. 
We consider the mapping relationship between the mixture signals and separated sources as a nonlinear transformation, and hence nonlinear models such as deep neural networks (DNNs) are desirable.

In this paper, we propose a general monaural source separation framework to jointly model all sources within a mixture as targets to a deep recurrent neural network (DRNN). We propose to utilize the constraints between the original mixture and the output predictions through time-frequency mask functions and jointly optimize the time-frequency functions along with the deep recurrent neural network. 
Given a mixture signal, the proposed approach directly reconstructs the predictions of target sources in an end-to-end fashion. In addition, given that there are predicted results of competing sources in the output layer, we further propose a discriminative training criterion for enhancing the source to interference ratio. 
We extend our previous work in \cite{Huang_DNN_Separation_ICASSP2014} and \cite{Huang_DRNN_ISMIR2014} and propose a general framework for monaural source separation tasks with applications to speech separation, singing voice separation, and speech denoising. 
We further extend our speech separation experiments in \cite{Huang_DNN_Separation_ICASSP2014} to a 
larger speech corpus, the TSP dataset \cite{kabal2002tsp},
with different model architectures and different speaker genders, and we extend our proposed framework to speech denoising tasks under various matched and mismatched conditions.

The organization of this paper is as follows:
Section \ref{sec:related_work} reviews and compares recent monaural source separation work based on deep learning models. Section \ref{sec:proposed_method} introduces the proposed methods, including the deep recurrent neural networks, joint optimization of deep learning models and soft time-frequency masking functions, and the training objectives. Section \ref{sec:exp} presents the experimental setting and results using the TSP \cite{kabal2002tsp}, MIR-1K \cite{Chao-Ling_Jang_2010}, and TIMIT \cite{garofolo1993timit} datasets for speech separation, singing voice separation, and speech denoising tasks, respectively. We conclude the paper in Section \ref{sec:conclusion}.

\section{Related Work}
\label{sec:related_work}
Recently, deep learning based methods have started to attract much attention in the source separation research community by modeling the nonlinear mapping relationship between mixture and separated signals.
Prior work on deep learning based source separation can be categorized into three categories, depending on the interaction between input mixture and output targets. 

{\bf Denoising-based approaches}:
These methods utilize deep learning based models to learn the mapping from the mixture signals to one of the sources among the mixture signals. 
In the speech recognition task, given noisy features, Maas et al. \cite{maas2012recurrent} proposed to apply a DRNN to predict clean speech features. In the speech enhancement task, Xu et al. \cite{Xu_RegreesionDNN_2014} and Liu et al. \cite{Liu_denoising_interspeech14} proposed to use a DNN for predicting clean speech signals given noisy speech signals. 
The denoising methods do not consider the relationships between target and other sources in the mixture, which is suboptimal in the source separation framework where all the sources are important.
In contrast, our proposed model considers \textit{all} sources in the mixture and utilizes the relationship among the sources to formulate time-frequency masks.

{\bf Time-frequency mask based approaches}:
A time-frequency mask \cite{wang2008time} considers the relationships among the sources in a mixture signal, enforces the constraints between an input mixture and the output predictions, and hence results in smooth prediction results. 
Weninger et al. \cite{Weninger_LSTMseparation_ICASSP14} trained two long short-term memory (LSTM) RNNs for predicting speech and noise, respectively. A final prediction is made by applying a time-frequency mask based on the speech and noise predictions.
Instead of training a model for each source and applying the time-frequency mask separately, our proposed model jointly optimizes time-frequency masks with a network which models all the sources directly.

Another type of approach is to apply deep learning models to predict a time-frequency mask for one of the sources. After the time-frequency mask is learned, the estimated source is obtained by multiplying the learned time-frequency mask with an input mixture.
Nie et al. \cite{Nie_DSNTS_2014} utilized deep stacking networks with time series inputs and a re-threshold method to predict an ideal binary mask. 
Narayanan and Wang \cite{narayananideal} and Wang and Wang \cite{Wang_Speech_Separation2013} proposed a two-stage framework (DNNs with a one-layer perceptron and DNNs with an SVM) for predicting a time-frequency mask. 
Wang et al. \cite{Wang_Wang_target_2014} recently proposed to train deep neural networks for different targets, including ideal ratio mask, FFT-mask, and Gammatone frequency power spectrum for speech separation tasks.
Our proposed approach learns time-frequency masks for all the sources internally with the DRNNs and directly optimizes separated results with respect to ground truth signals in an end-to-end fashion.

{\bf Multiple-target based approaches:}
These methods model all output sources in a mixture as deep learning model training targets. 
Tu et al. \cite{Tu_ICSP14} proposed modeling clean speech and noise as the output targets for a robust ASR task. 
However, the authors do not consider the constraint that the sum of all the sources is the original mixture. 
Grais et al. \cite{Grais_single_channel_2014} proposed using a deep neural network to predict two scores corresponding to the probabilities of two different sources respectively given a frame of normalized magnitude spectrum. 
Our proposed method also models all sources as training targets. We further enforce the constraints between an input mixture and the output predictions through time-frequency masks which are learned along with DRNNs.



\begin{figure*}[ht!]
	\centering
	\subfigure[1-layer RNN]{
		\includegraphics[height=2.in]{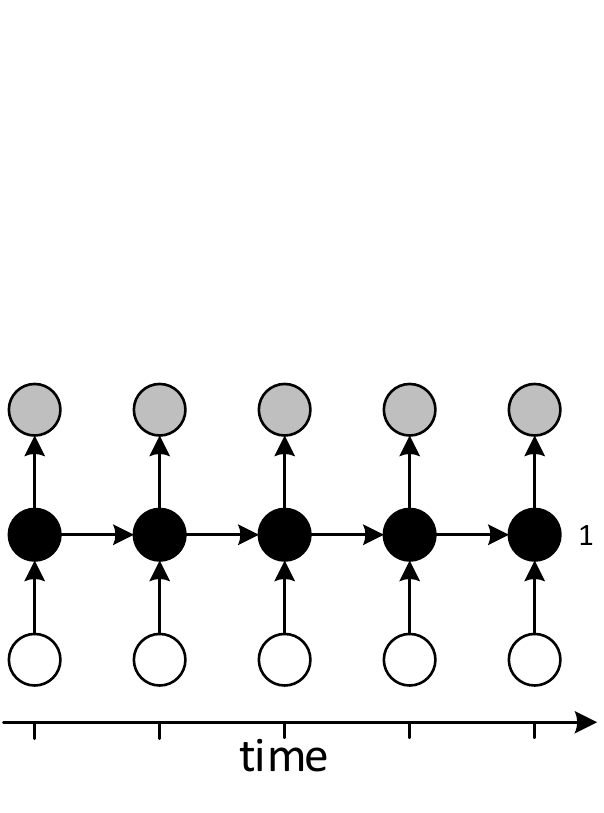}
	}
	\subfigure[L-layer DRNN-$l$]{
		\includegraphics[height=2.in]{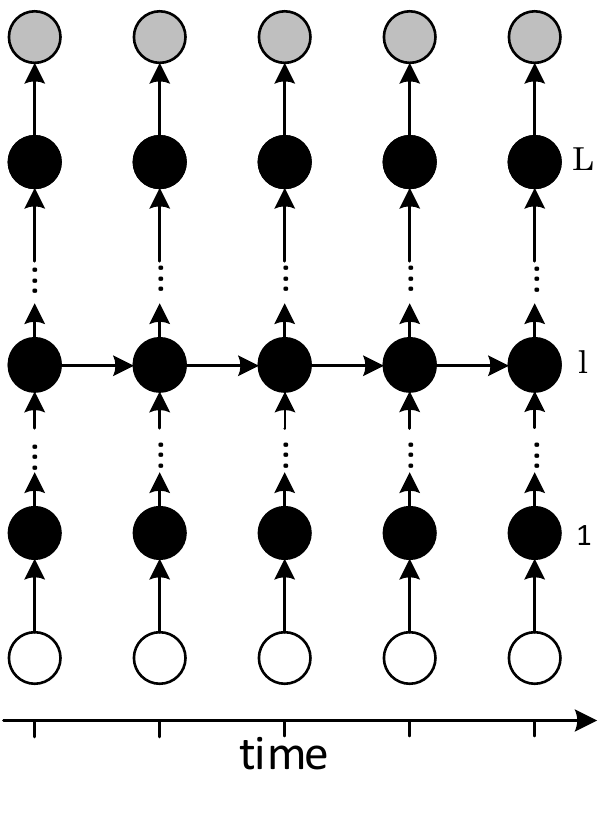}
	}
	\subfigure[L-layer stacked RNN (sRNN)]{
		\includegraphics[height=2.in]{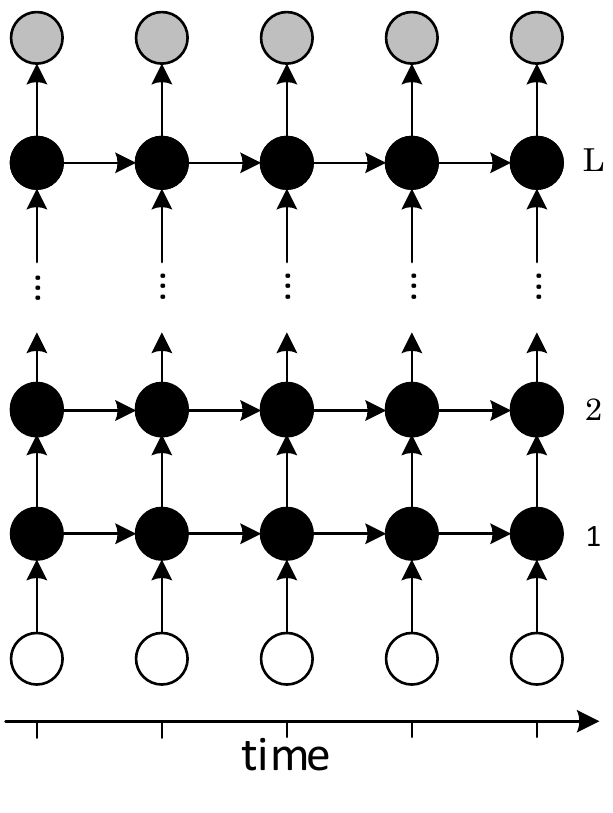}
	}
	\vspace{-2mm}	
		\caption{{Deep Recurrent Neural Network (DRNN) architectures: Arrows represent connection matrices. Black, white, and gray circles represent input frames, hidden states, and output frames, respectively. The architecture in (a) is a standard recurrent neural network, (b) is an $L$ hidden layer DRNN with recurrent connection at the $l$-th layer (denoted by DRNN-$l$), and (c) is an $L$ hidden layer DRNN with recurrent connections at all levels (denoted by stacked RNN).}}
	\label{fig:DRNN_architectures}
	\vspace{-1mm}	
\end{figure*}

\section{Proposed Methods}
\label{sec:proposed_method}
\subsection{Deep Recurrent Neural Networks}

Given that audio signals are time series in nature, we propose to model the temporal information using deep recurrent neural networks for monaural source separation tasks. 
To capture the contextual information among audio signals, one way is to concatenate neighboring audio features, e.g., magnitude spectra, together as input features to a deep neural network. However, the number of neural network parameters increases proportionally to the input dimension and the number of neighbors in time. Hence, the size of the concatenating window is limited. 
Another approach is to utilize recurrent neural networks (RNNs) for modeling the temporal information. 
An RNN can be considered as a DNN with indefinitely many layers, which introduce the memory from previous time steps, as shown in Figure \ref{fig:DRNN_architectures} (a). 
The potential weakness for RNNs is that RNNs lack hierarchical processing of the input at the current time step. To further provide the hierarchical information through multiple time scales, deep recurrent neural networks (DRNNs) are explored \cite{hermans2013training, Pascanu_iclr2014}. 
We formulate DRNNs in two schemes as shown in Figure \ref{fig:DRNN_architectures} (b) and Figure \ref{fig:DRNN_architectures} (c). 
The Figure \ref{fig:DRNN_architectures} (b) is an $L$ hidden layer DRNN with temporal connection at the $l$-th layer. The Figure \ref{fig:DRNN_architectures} (c) is an $L$ hidden layer DRNN with full temporal connections (called stacked RNN (sRNN) in \cite{Pascanu_iclr2014}).
Formally, we define the two DRNN schemes as follows. 
Suppose there is an $L$ hidden layer DRNN with the recurrent connection at the $l$-th layer, the $l$-th hidden activation at time $t$, $\mathbf{h}_{t}^l$, is defined as:
\begin{align*}
\mathbf{h}^{l}_t&=f_\mathrm{h}(\mathbf{x}_t, \mathbf{h}^{l}_{t-1})\\
	   &= \phi_l\left(\mathbf{U}^{l} \mathbf{h}^l_{t-1} + \mathbf{W}^{l} \phi_{l-1} \left(\mathbf{W}^{l-1} \left(\ldots \phi_1\left(\mathbf{W}^{1} \mathbf{x}_t \right) \right) \right) \right)
\numberthis 
\end{align*}
and the output $\mathbf{y}_t$ is defined as: 
\begin{align*}
\mathbf{y}_t&= f_\mathrm{o}(\mathbf{h}^{l}_t)\\
   &= 
       \mathbf{W}^{L}	\phi_{L-1} \left(\mathbf{W}^{L-1} \left(\ldots \phi_l\left( 
	\mathbf{W}^{l} \mathbf{h}^l_t \right) \right) \right) 
\numberthis 
\end{align*}
where $f_\mathrm{h}$ and $f_\mathrm{o}$ are a state transition function and an output function, respectively, $\mathbf{x}_t$ is the input to the network at time $t$, $\phi_l(\cdot)$ is an element-wise nonlinear function at the $l$-th layer, $\mathbf{W}^l$ is the weight matrix for the $l$-th layer, and $\mathbf{U}^l$ is the weight matrix for the recurrent connection at the $l$-th layer. The recurrent weight matrix $\mathbf{U}^k$ is a zero matrix for the rest of the layers where $k\neq l$. The output layer is a linear layer. 

The stacked RNNs, as shown in Figure \ref{fig:DRNN_architectures} (c), have multiple levels of transition functions, defined as:
\begin{align*}
\mathbf{h}^{l}_t&=f_\mathrm{h}(\mathbf{h}^{l-1}_t, \mathbf{h}^l_{t-1})\\
	&=\phi_l (\mathbf{U}^{l} \mathbf{h}^l_{t-1} +	\mathbf{W}^{l} \mathbf{h}^{l-1}_t)
	\numberthis
	\label{eq:hidden_activation} 
\end{align*}
where $\mathbf{h}^l_t$ is the hidden state of the $l$-th layer at time $t$, $\phi_l(\cdot)$ is an element-wise nonlinear function at the $l$-th layer, $\mathbf{W}^l$ is the weight matrix for the $l$-th layer, and $\mathbf{U}^l$ is the weight matrix for the recurrent connection at the $l$-th layer. When the layer $l=1$, the hidden activation $\mathbf{h}^{1}_t$ is computed using $\mathbf{h}^{0}_{t} = \mathbf{x}_{t}$.
For the nonlinear function $\phi_l(\cdot)$, similar to \cite{Glorot+al-AI-2011}, we empirically found that using the rectified linear unit $\phi_l(\mathbf{x})=\max(\mathbf{0},\mathbf{x})$ performs better compared to using a sigmoid or tanh function in our experiments. Note that a DNN can be regarded as a DRNN with the temporal weight matrix $\mathbf{U}^{l}$ as a zero matrix. 

For the computation complexity, given the same input features, during the forward-propagation stage, 
a DRNN with $L$ hidden layers, $m$ hidden units, and a temporal connection at the $l$-th layer requires an extra $\Theta(m^2)$ IEEE floating point storage buffer to store the temporal weight matrix $\mathbf{U}^{l}$, and extra $\Theta(m^2)$ multiply-add operations to compute the hidden activations in Eq. \eqref{eq:hidden_activation} at the $l$-th layer, compared to a DNN with $L$ hidden layers and $m$ hidden units.
During the back-propagation stage, DRNN uses back-propagation through time (BPTT) \cite{mozer1989focused, werbos1990backpropagation} to update network parameters. 
Given an input sequence with $T$ time steps in length, the DRNN with an $l$-th layer temporal connection requires an extra $\Theta(Tm)$ space to keep hidden activations in memory and requires 
$\Theta(Tm^2)$ operations ($\Theta(m^2)$ operations per time step) for updating parameters, compared to a DNN \cite{Williams:1995:GLA:201784.201801}.
Indeed, the only pragmatically significant computational cost of a DRNN with respect to a DNN is that the recurrent layer limits the granularity with which back-propagation can be parallelized. 
As gradient updates based on sequential steps cannot be computed in parallel, for improving the efficiency of DRNN training, utterances are chopped into sequences of at most 100 time steps.


\subsection{Model Architecture}
\label{sec: time_freq_mask}

We consider the setting where there are two sources additively mixed together, though our proposed framework can be generalized to more than two sources. 
At time $t$, the training input $\mathbf{x}_t$ of the network is the concatenation of features, e.g., logmel features or magnitude spectra, from a mixture within a window. 
The output targets $\mathbf{y}_{\mathbf{1}_t}\in \mathbb{R}^F$ and $\mathbf{y}_{\mathbf{2}_t}\in \mathbb{R}^F$ and the output predictions $\mathbf{\hat{y}}_{\mathbf{1}_t}\in\mathbb{R}^F$ and $\mathbf{\hat{y}}_{\mathbf{2}_t}\in \mathbb{R}^F$ of the deep learning models are the magnitude spectra of different sources, where $F$ is the magnitude spectral dimension.

\begin{figure}[t]
\includegraphics[width=2.95in]{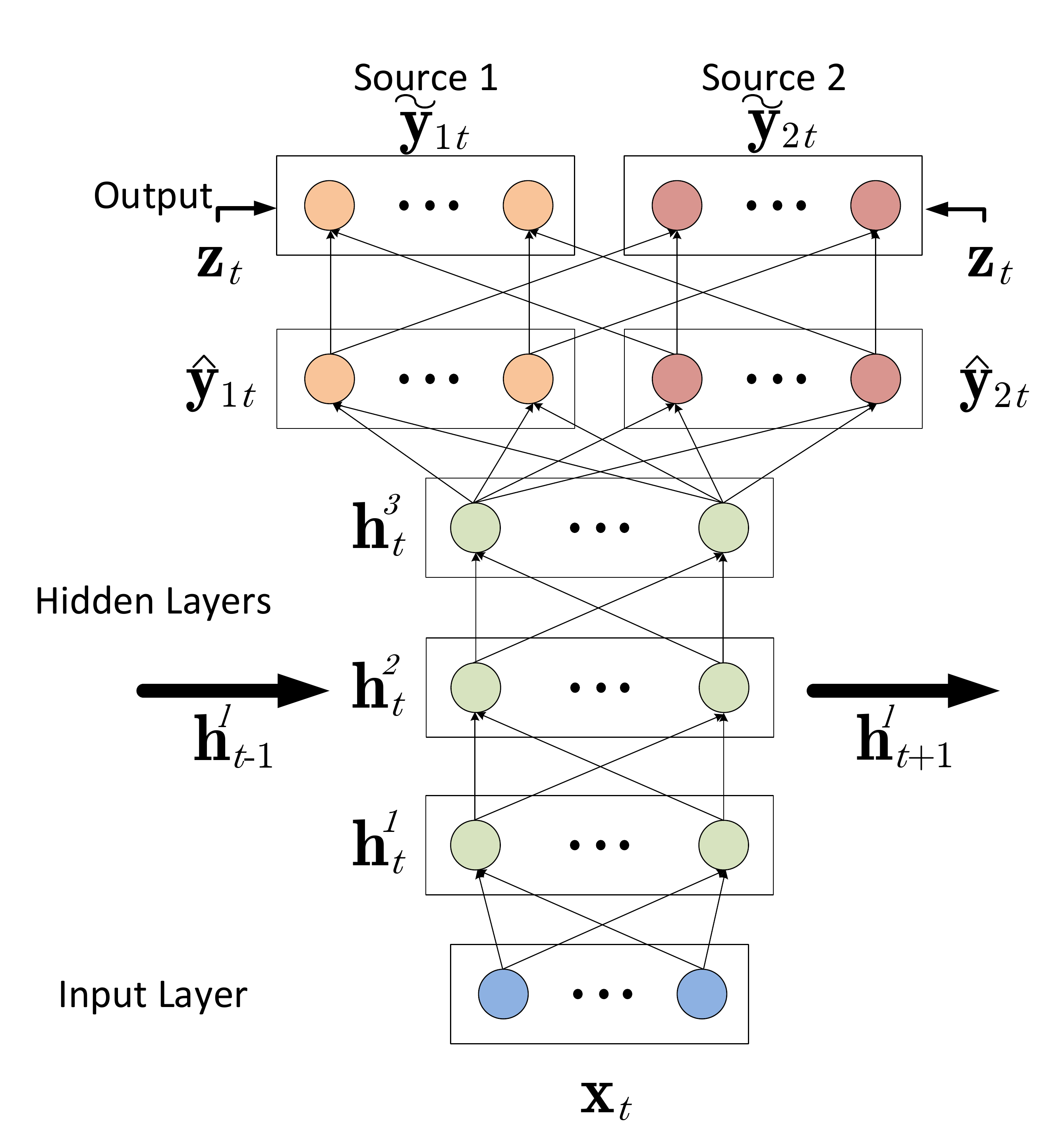}
\vspace{-2mm}
\caption{{Proposed neural network architecture, which can be viewed as the $t$-th column in Figure \ref{fig:DRNN_architectures}. We propose to jointly optimize time-frequency masking functions as a layer with a deep recurrent neural network.}}
\label{fig:network}
\end{figure}

Since our goal is to separate different sources from a mixture, instead of learning one of the sources as the target, we propose to simultaneously model all the sources. 
Figure \ref{fig:network} shows an example of the architecture, which can be viewed as the $t$-th column in Figure \ref{fig:DRNN_architectures}.

Moreover, we find it useful to further smooth the source separation results with a time-frequency masking technique, for example, binary time-frequency masking or soft time-frequency masking \cite{Yilmaz_TF_masking_2004, wang2008time, Huang_RPCA_Separation_ICASSP2012,Huang_DNN_Separation_ICASSP2014}.
The time-frequency masking function enforces the constraint that the sum of the prediction results is equal to the original mixture.
Given the input features $\mathbf{x}_{t}$ from the mixture, we obtain the output predictions 
$\mathbf{\hat{y}}_{\mathbf{1}_{t}}$ and $\mathbf{\hat{y}}_{\mathbf{2}_{t}}$ through the network. 
The soft time-frequency mask $\mathbf{m}_t\in\mathbb{R}^F$ is defined as follows:
\begin{equation}
\label{eq:soft_mask}
\mathbf{m}_t = \dfrac{|\mathbf{\hat{y}}_{\mathbf{1}_t}|}{|\mathbf{\hat{y}}_{\mathbf{1}_t}|+ |\mathbf{\hat{y}}_{\mathbf{2}_t}|}
\end{equation}
where the addition and division operators are element-wise operations.

Similar to \cite{Weninger_LSTMseparation_ICASSP14}, a standard approach is to apply the time-frequency masks $\mathbf{m}_t$  and $\mathbf{1}-\mathbf{m}_t$ to the magnitude spectra $\mathbf{z}_{t} \in\mathbb{R}^F$ of the mixture signals, and obtain the estimated separation spectra $\mathbf{\hat{s}}_{\mathbf{1}_t}\in\mathbb{R}^F$ and $\mathbf{\hat{s}}_{\mathbf{2}_t}\in\mathbb{R}^F$, which correspond to sources 1 and 2, as follows: 
\vspace{-.5mm}

\begin{equation}
\begin{array}{r@{}l}
\label{eq:applied_mask}
   \mathbf{\hat{s}}_{\mathbf{1}_t}&{}=  \mathbf{m}_t \odot \mathbf{z}_{t}\\ 
   \mathbf{\hat{s}}_{\mathbf{2}_t} &{} = \left(\mathbf{1}-\mathbf{m}_t \right)\odot\mathbf{z}_{t}
\end{array}
\end{equation}
where the subtraction and $\odot$ (Hadamard product) operators are element-wise operations.

Given the benefit of smoothing separation and enforcing the constraints between an input mixture and the output predictions using time-frequency masks, we propose to incorporate the time-frequency masking functions as a layer in the neural network. Instead of training the neural network and applying the time-frequency masks to the predictions separately, we propose to jointly train the deep learning model with the time-frequency masking functions.
We add an extra layer to the original output of the neural network as follows:

\begin{equation}
\begin{array}{r@{}l}
\label{eq:apply_soft_mask}
  \mathbf{\tilde{y}}_{\mathbf{1}_t} &{}= \dfrac{ |\mathbf{\hat{y}}_{\mathbf{1}_t}| }{|\mathbf{\hat{y}}_{\mathbf{1}_t}|+|\mathbf{\hat{y}}_{\mathbf{2}_t}|} \odot\mathbf{z}_t\\
  \mathbf{\tilde{y}}_{\mathbf{2}_t} &{}= \dfrac{ |\mathbf{\hat{y}}_{\mathbf{2}_t}| }{|\mathbf{\hat{y}}_{\mathbf{1}_t}|+|\mathbf{\hat{y}}_{\mathbf{2}_t}|} \odot\mathbf{z}_t
\end{array}
\end{equation}
where the addition, division, and $\odot$ (Hadamard product) operators are element-wise operations. The architecture is shown in Figure \ref{fig:network}.
In this way, we can integrate the constraints into the network and optimize the network with the masking functions jointly. Note that although this extra layer is a deterministic layer, the network weights are optimized for the error metric between $\mathbf{\tilde{y}}_{\mathbf{1}_t}$, $\mathbf{\tilde{y}}_{\mathbf{2}_t}$ and $\mathbf{y}_{\mathbf{1}_t}$, $\mathbf{y}_{\mathbf{2}_t}$, using the back-propagation algorithm. 
The time domain signals are reconstructed based on the inverse short-time Fourier transform (ISTFT) of the estimated magnitude spectra along with the original mixture phase spectra.

\subsection{Training Objectives}

Given the output predictions $\mathbf{\tilde{y}}_{\mathbf{1}_{t}}$ and $\mathbf{\tilde{y}}_{\mathbf{2}_{t}}$ (or $\mathbf{\hat{y}}_{\mathbf{1}_{t}}$ and $\mathbf{\hat{y}}_{\mathbf{2}_{t}}$) 
of the original sources $\mathbf{y}_{\mathbf{1}_{t}}$ and $\mathbf{y}_{\mathbf{2}_{t}}$, $t=1, \ldots, T$, where $T$ is the length of an input sequence, we optimize the neural network parameters by minimizing the squared error:
\begin{equation}
\label{eq:mse}
J_{\mathrm{MSE}}=\frac{1}{2}\sum_{t=1}^T \left(\Vert\mathbf{\tilde{y}}_{\mathbf{1}_{t}}-\mathbf{y}_{\mathbf{1}_{t}}\Vert_{2}^2+\Vert\mathbf{\tilde{y}}_{\mathbf{2}_{t}}-\mathbf{y}_{\mathbf{2}_{t}}\Vert_{2}^2\right)
\end{equation}

\begin{figure*}[ht!]
	\centering
	\subfigure[Mixture]{
		\includegraphics[height=0.97in]{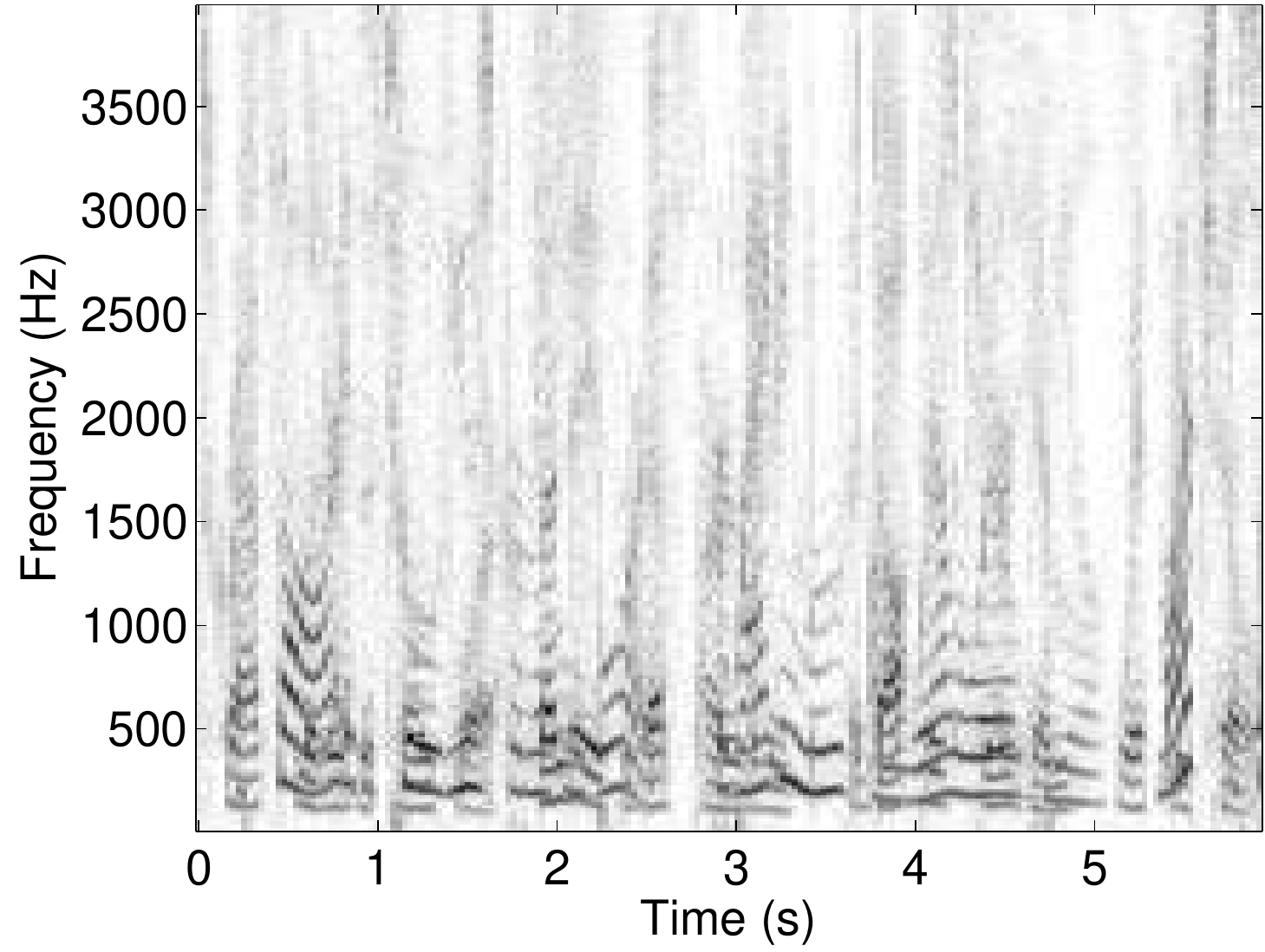}
	}
	\subfigure[Original female voice]{
		\includegraphics[height=0.97in]{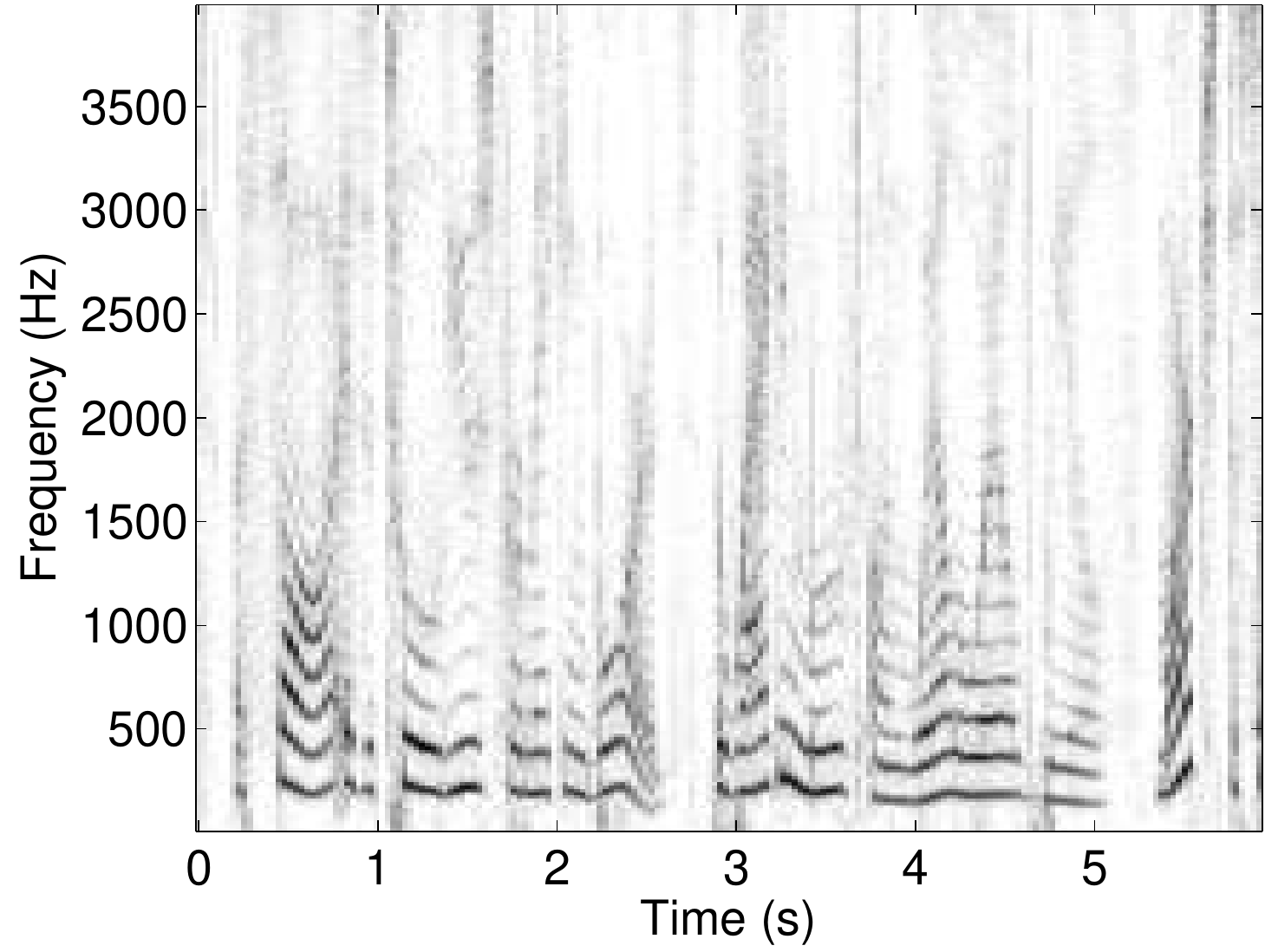}
	}
	\subfigure[Recovered female voice]{
		\includegraphics[height=0.97in]{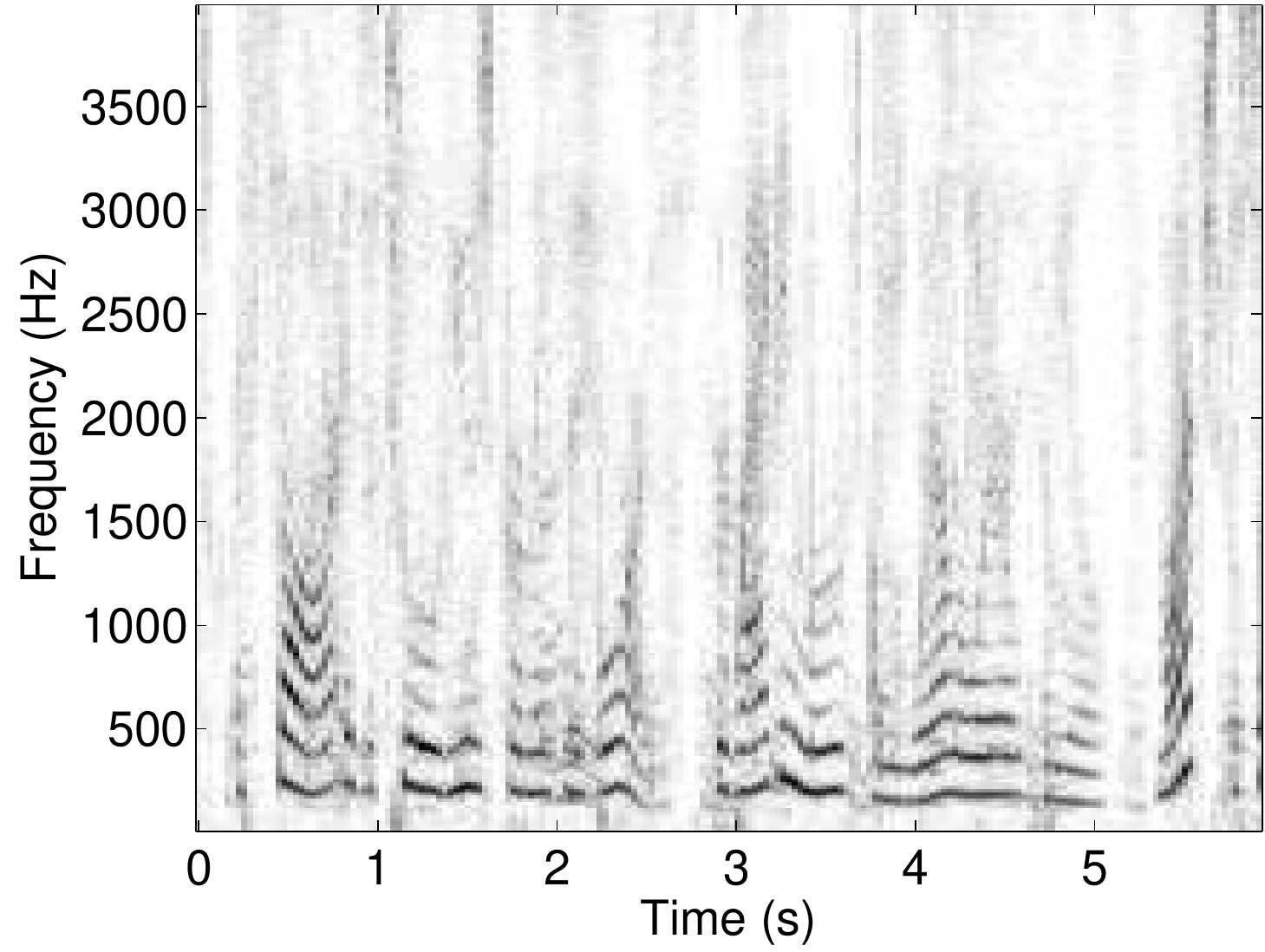}
	}
	\subfigure[Original male voice]{
		\includegraphics[height=0.97in]{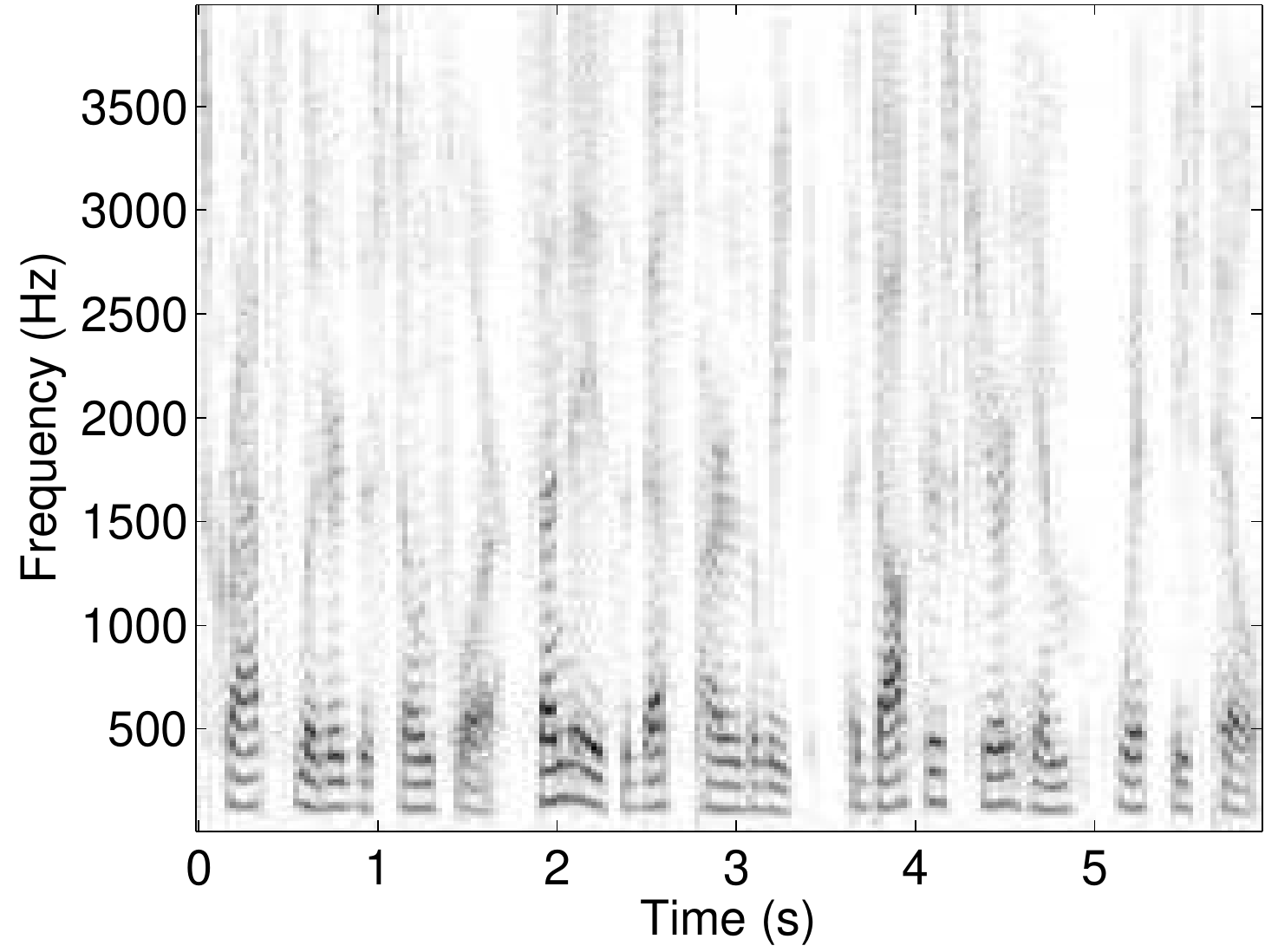}
	}
	\subfigure[Recovered male voice]{
		\includegraphics[height=0.97in]{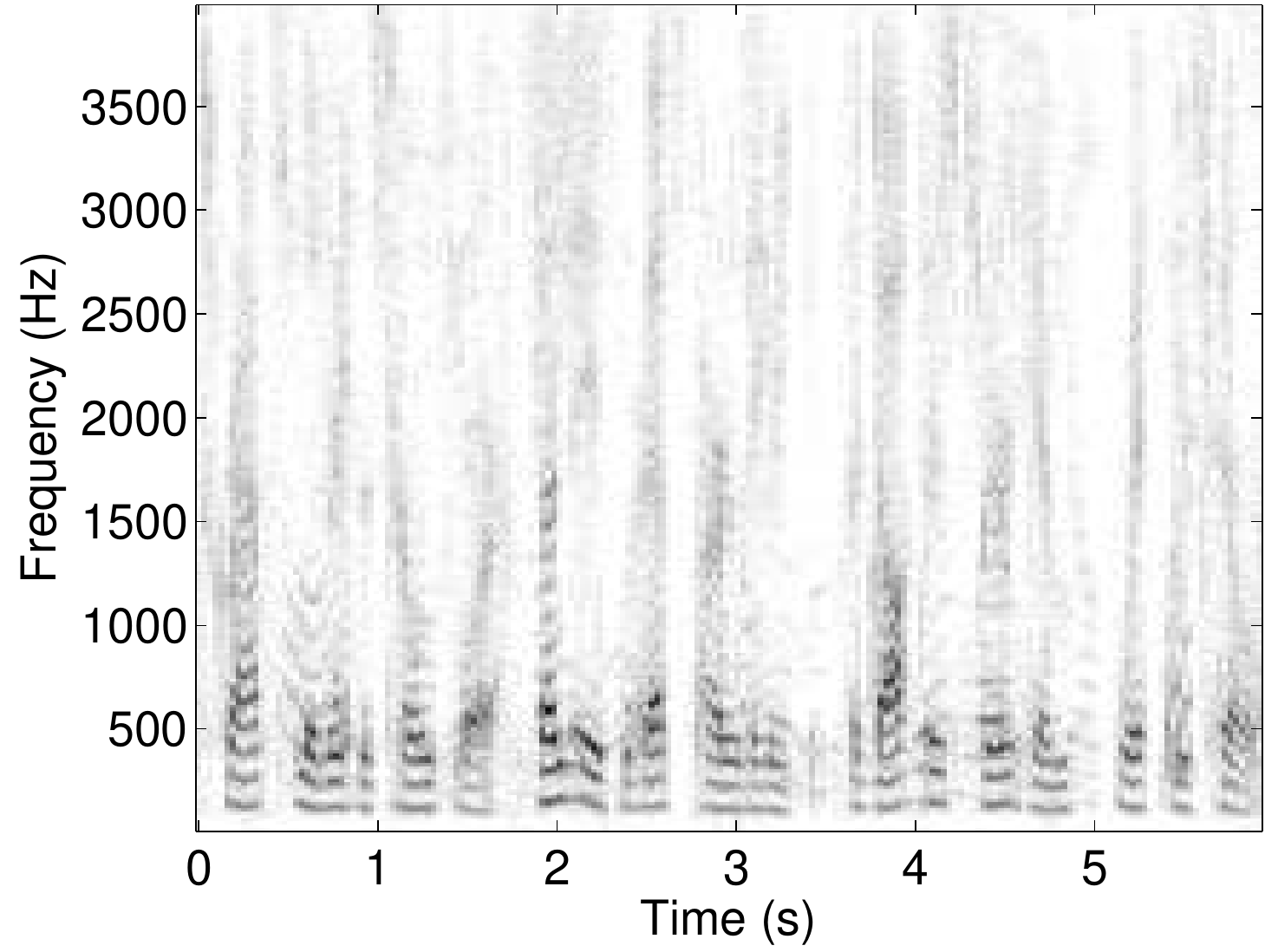}
	}
	\vspace{-4.mm}	
	\caption{{A speech separation example using the TSP dataset. (a) The mixture (female (FA) and male (MC) speech) magnitude spectrogram for a test clip in TSP; (b) the ground truth spectrogram of the female speech; (c) the separated female speech spectrogram from our proposed model (DRNN-1 + discrim); (d) the ground truth spectrogram of the male speech; (e) the separated male speech spectrogram from our proposed model (DRNN-1 + discrim).}} 
	\label{fig:TSP_separation_example}
	\vspace{-2mm}	
\end{figure*}

\begin{figure*}[ht!]
	\centering
	\subfigure[Mixture]{
		\includegraphics[height=0.97in]{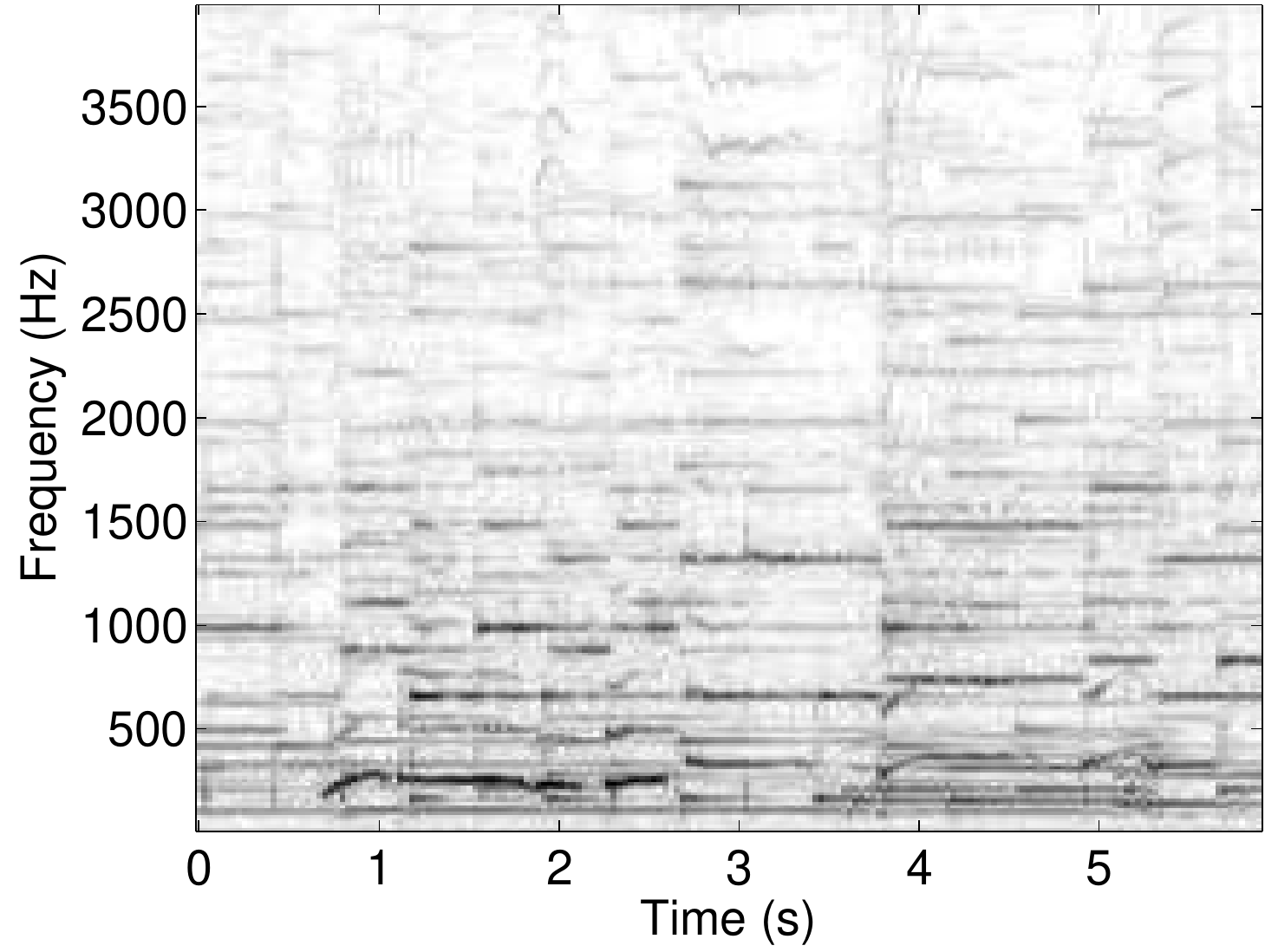}
	}
	\subfigure[Original singing]{
		\includegraphics[height=0.97in]{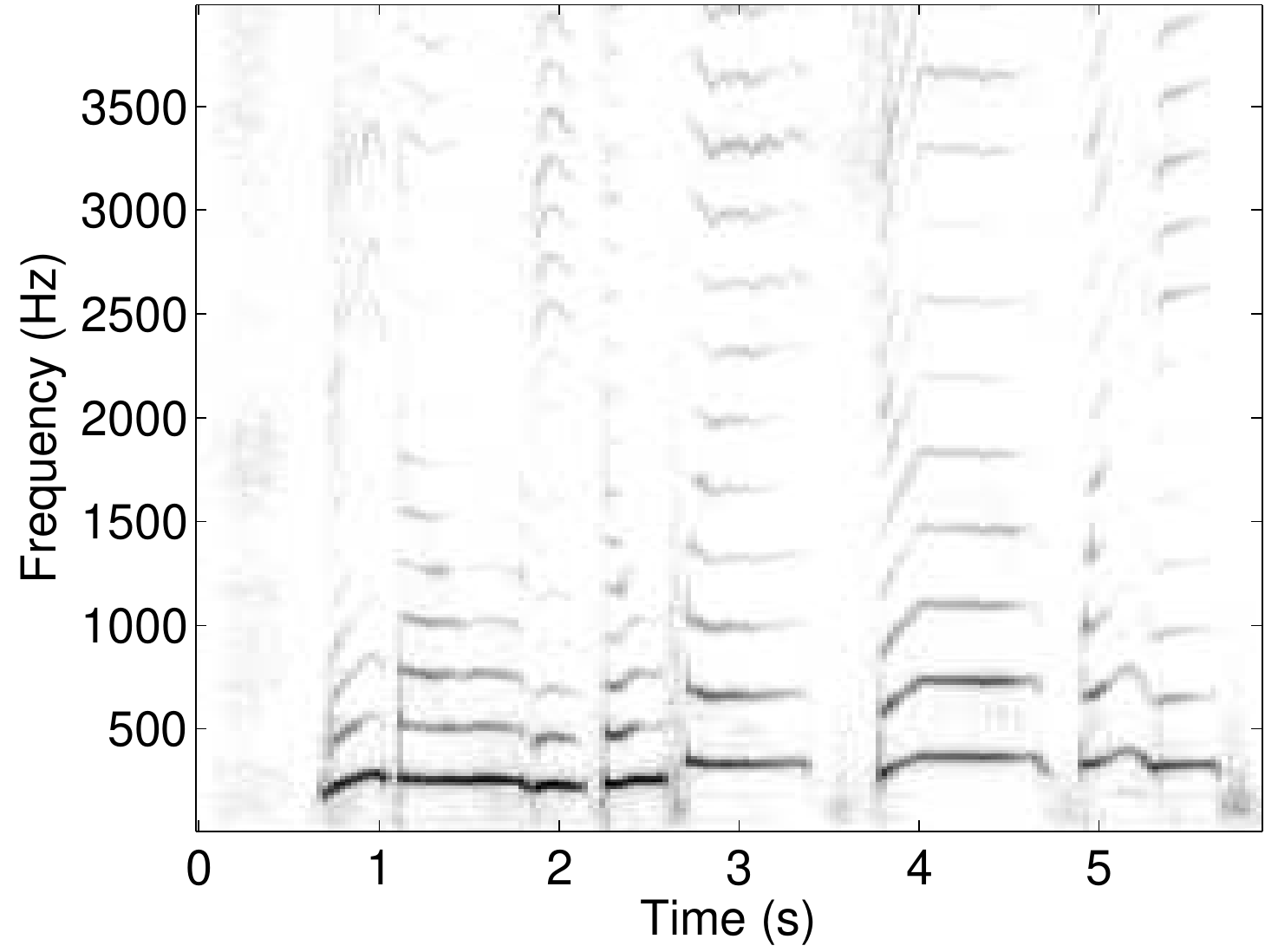}
	}
	\subfigure[Recovered singing]{
		\includegraphics[height=0.97in]{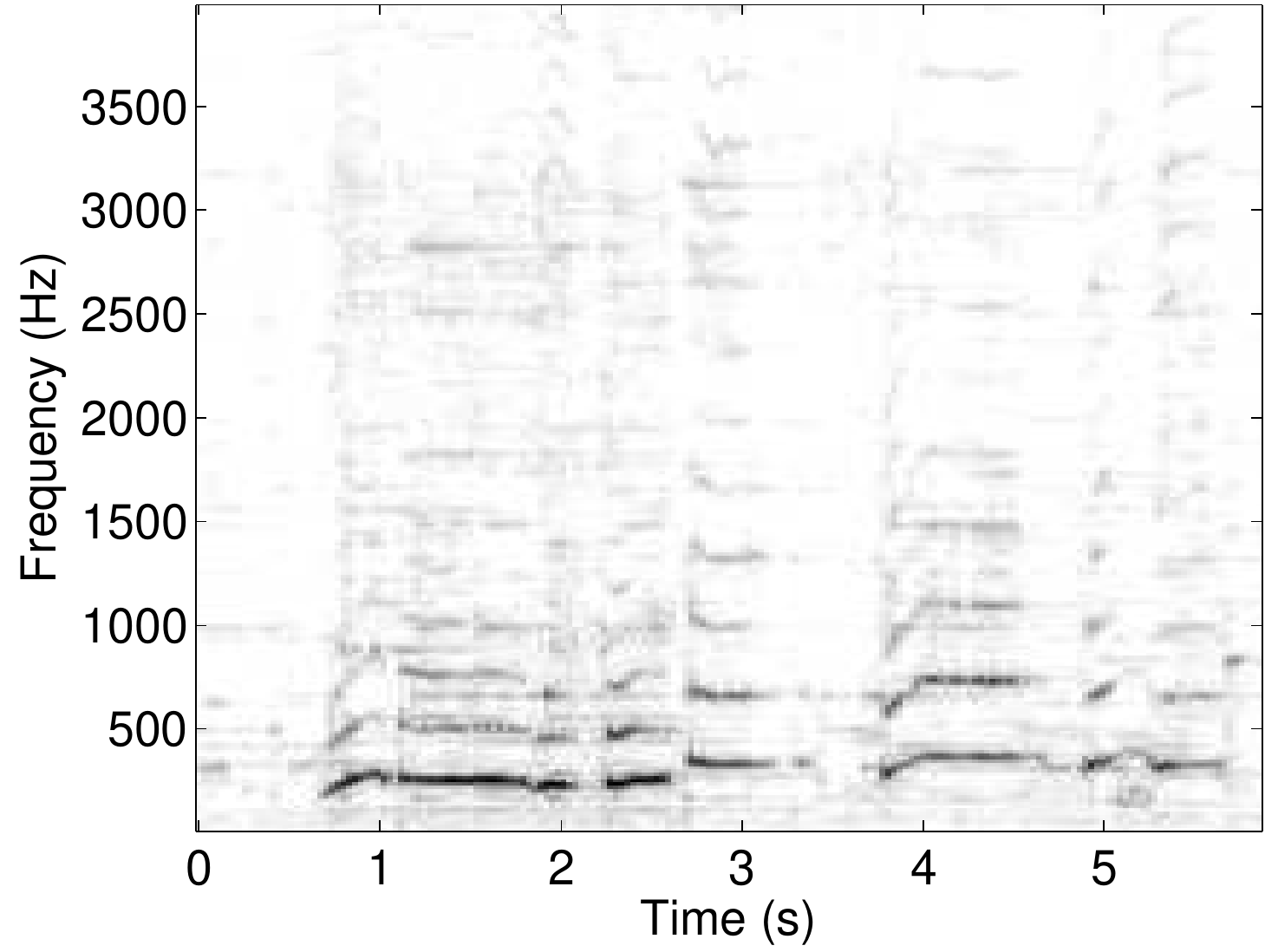}
	}
	\subfigure[Original music]{
		\includegraphics[height=0.97in]{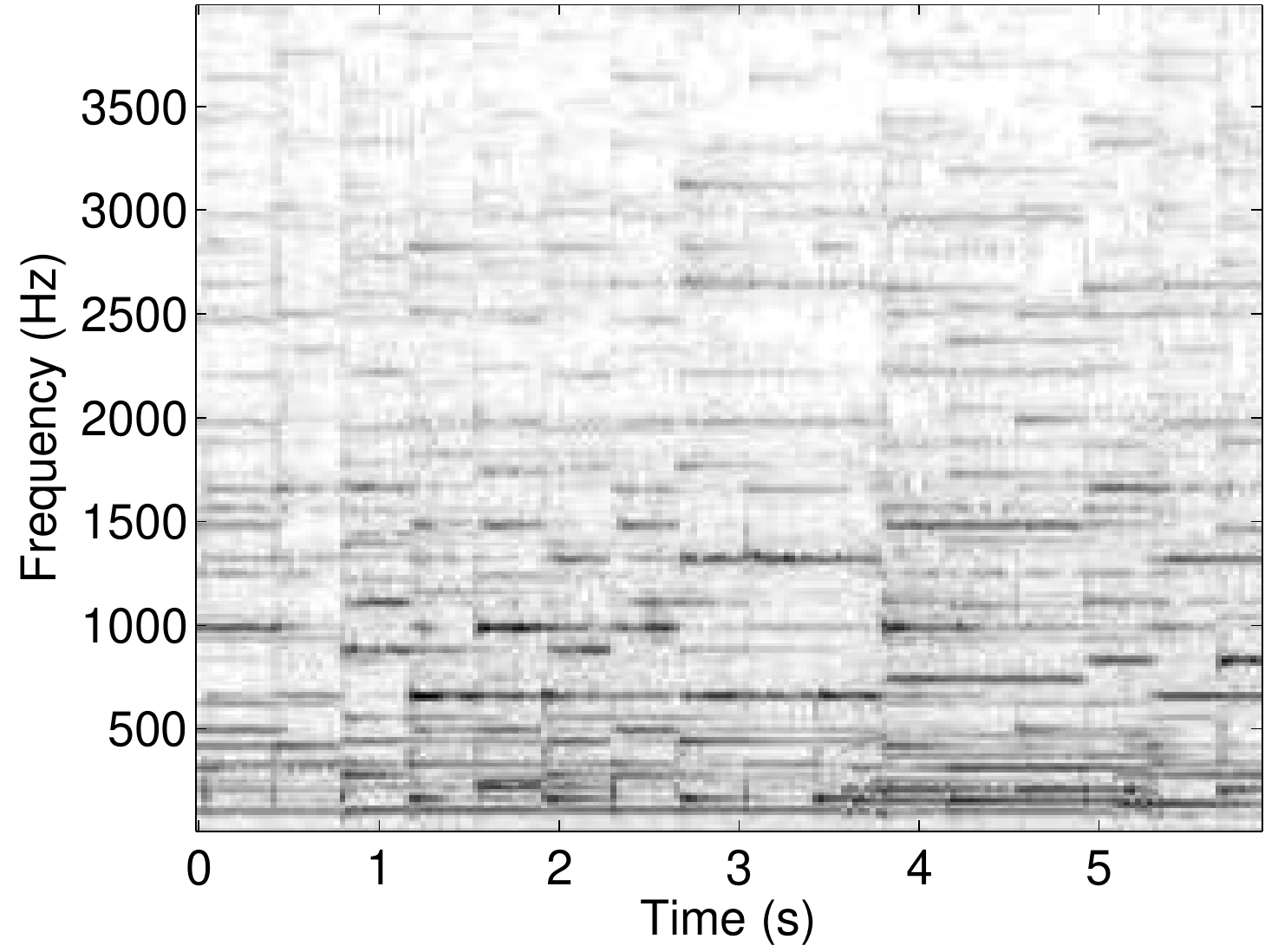}
	}
	\subfigure[Recovered music]{
		\includegraphics[height=0.97in]{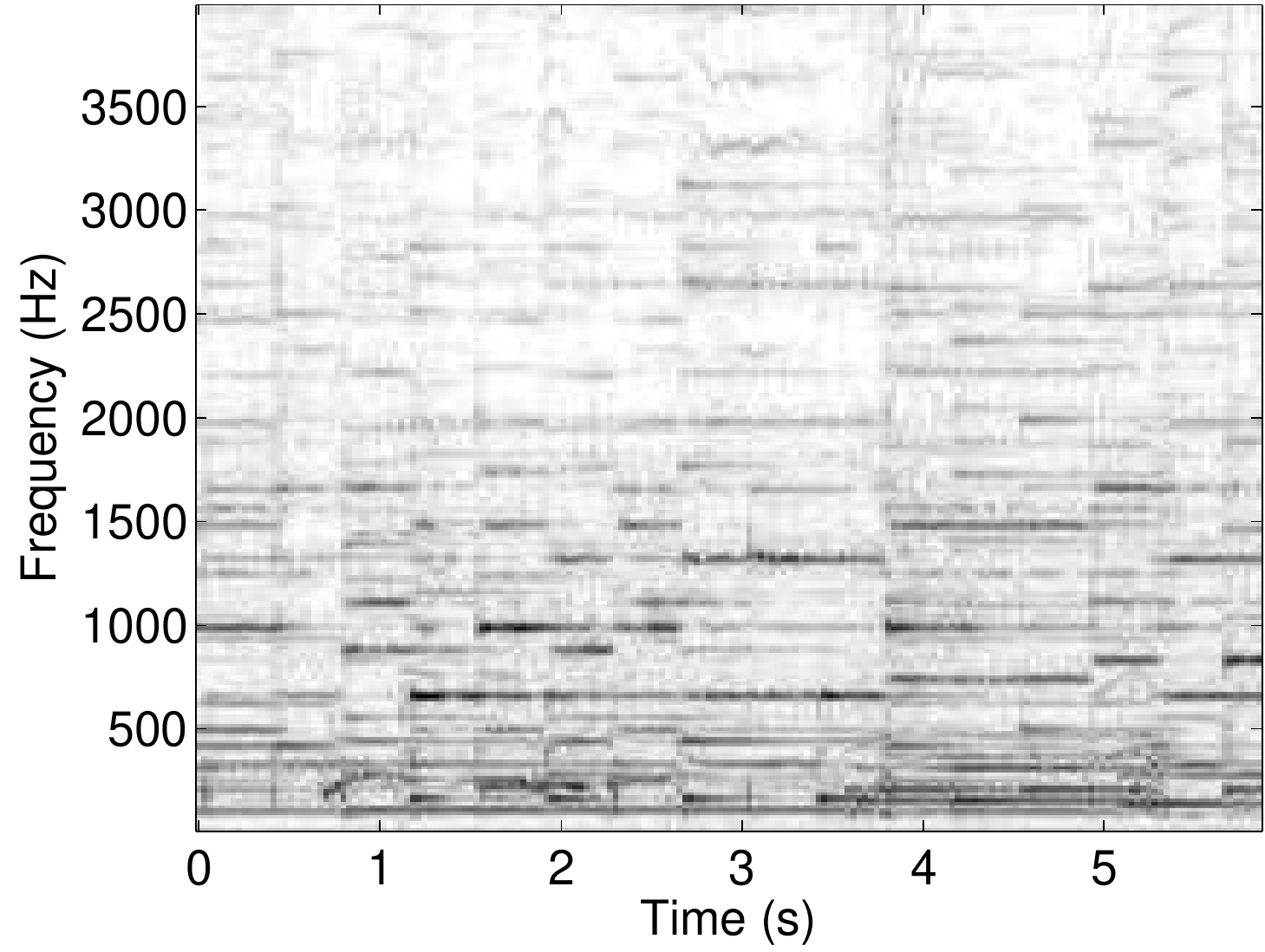}
	}
	\vspace{-4.mm}	
	\caption{{A singing voice separation example using the MIR-1K dataset. (a) The mixture (singing voice and music accompaniment) magnitude spectrogram for the clip Yifen\_2\_07 in MIR-1K; (b) the ground truth spectrogram for the singing voice; (c) the separated signing voice spectrogram from our proposed model (DRNN-2 + discrim); (d) the ground truth spectrogram for the music accompaniment; (e) the separated music accompaniment spectrogram from our proposed model (DRNN-2 + discrim).}} 
	\label{fig:mir_separation_example}
	\vspace{-2mm}	
\end{figure*}

\begin{figure*}[ht!]
	\centering
	\subfigure[Mixture]{
		\includegraphics[height=0.97in]{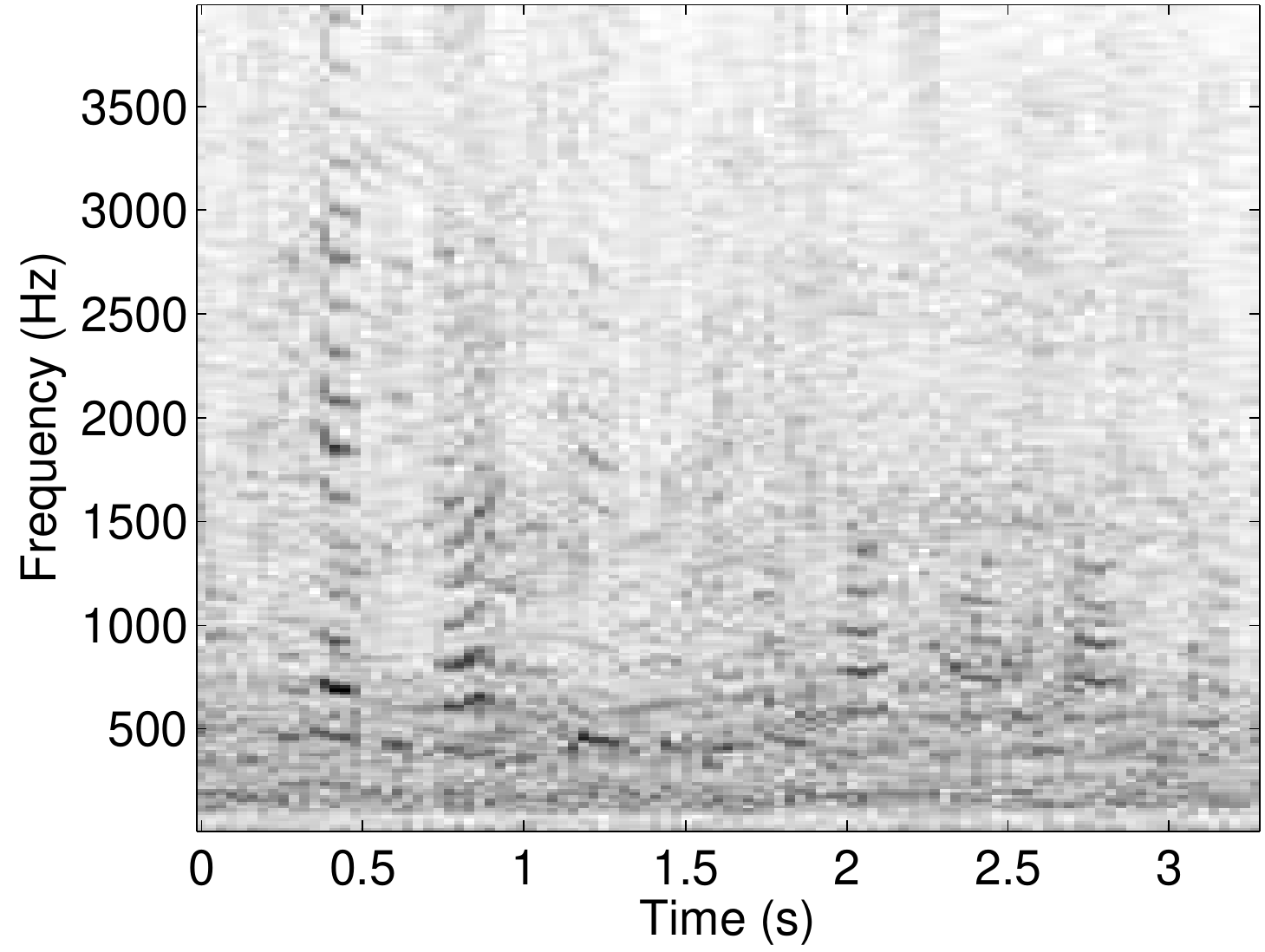}
	}
	\subfigure[Original speech]{
		\includegraphics[height=0.97in]{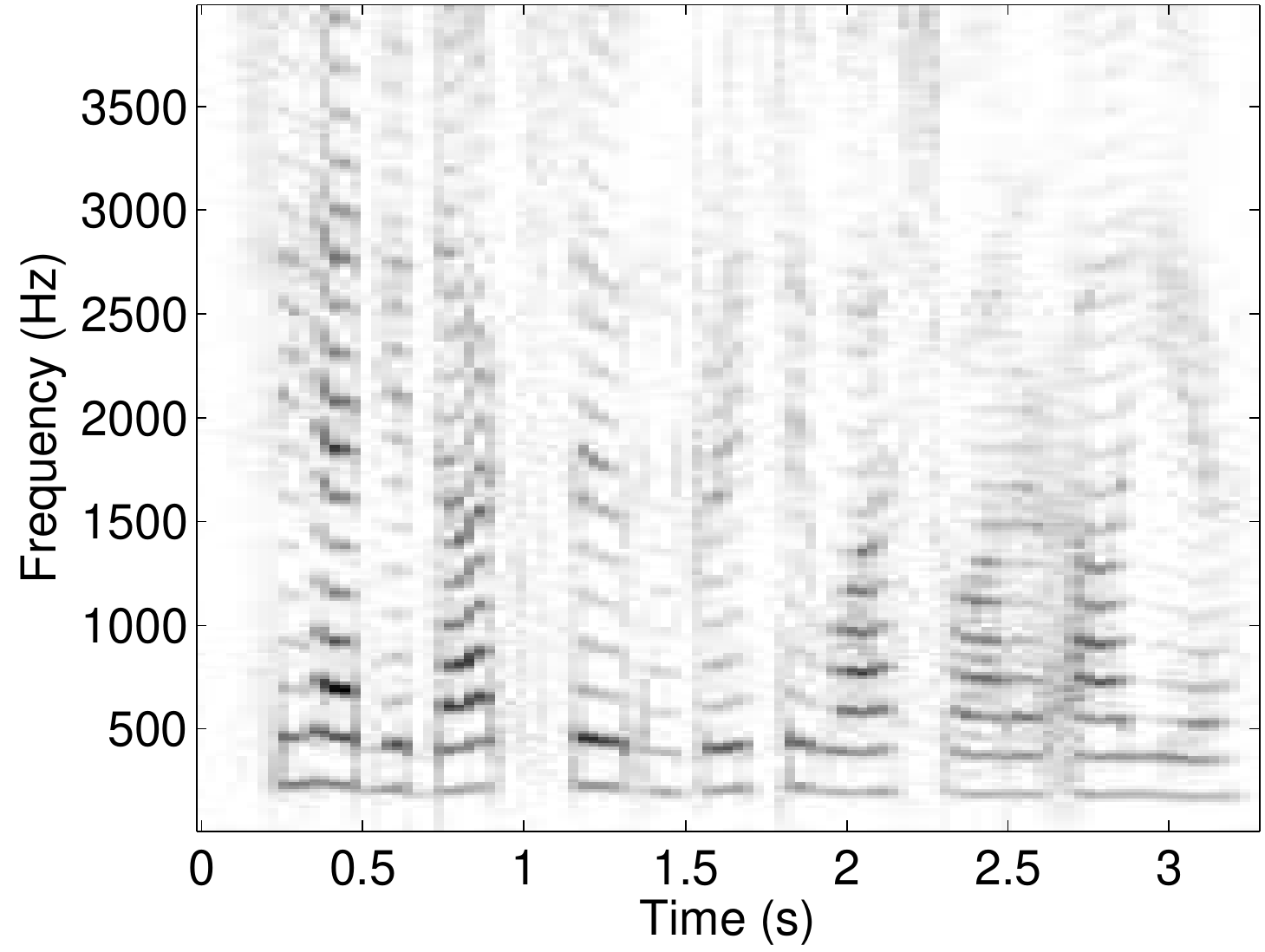}
	}
	\subfigure[Recovered speech]{
		\includegraphics[height=0.97in]{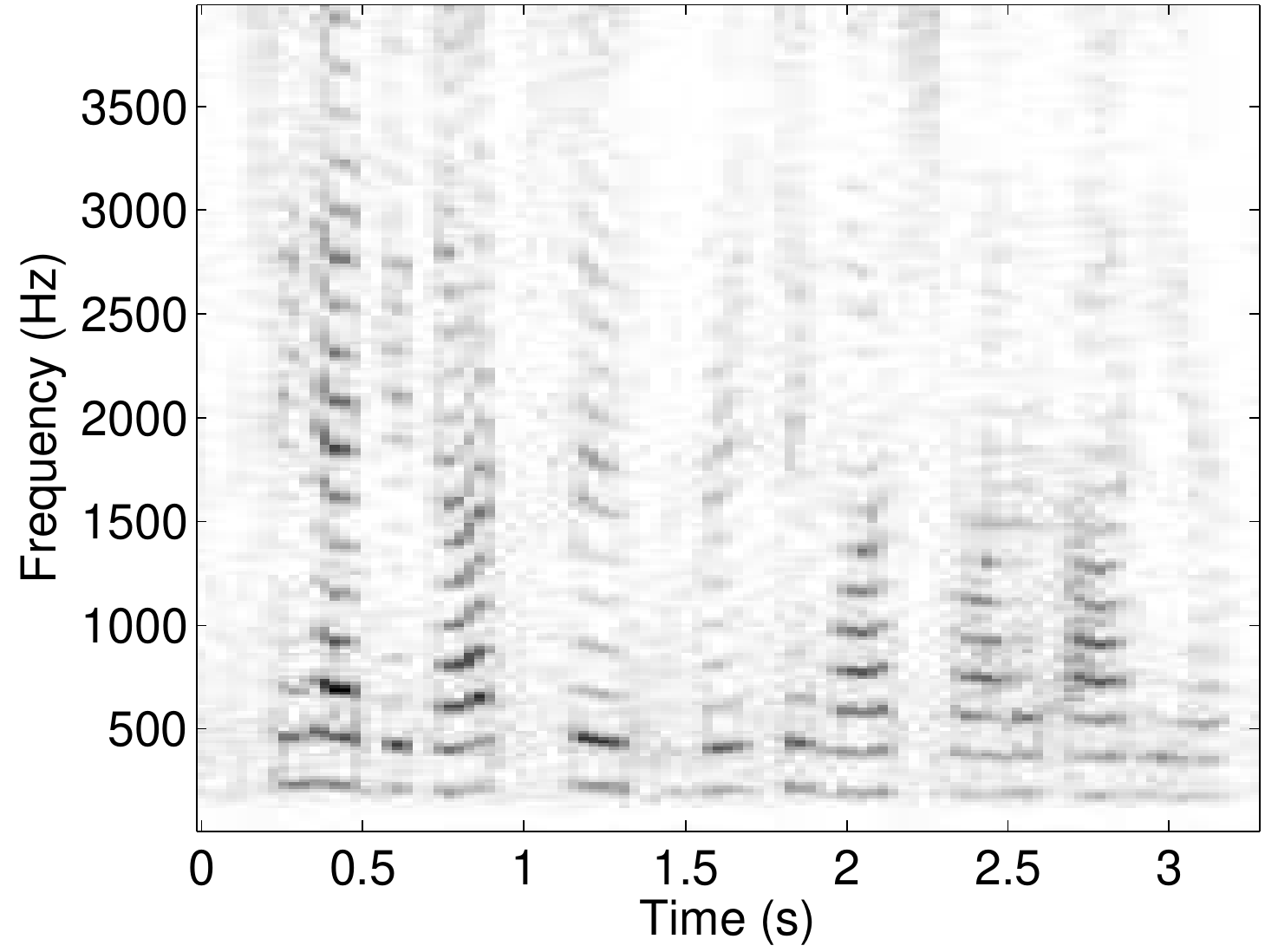}
	}
	\subfigure[Original noise]{
		\includegraphics[height=0.97in]{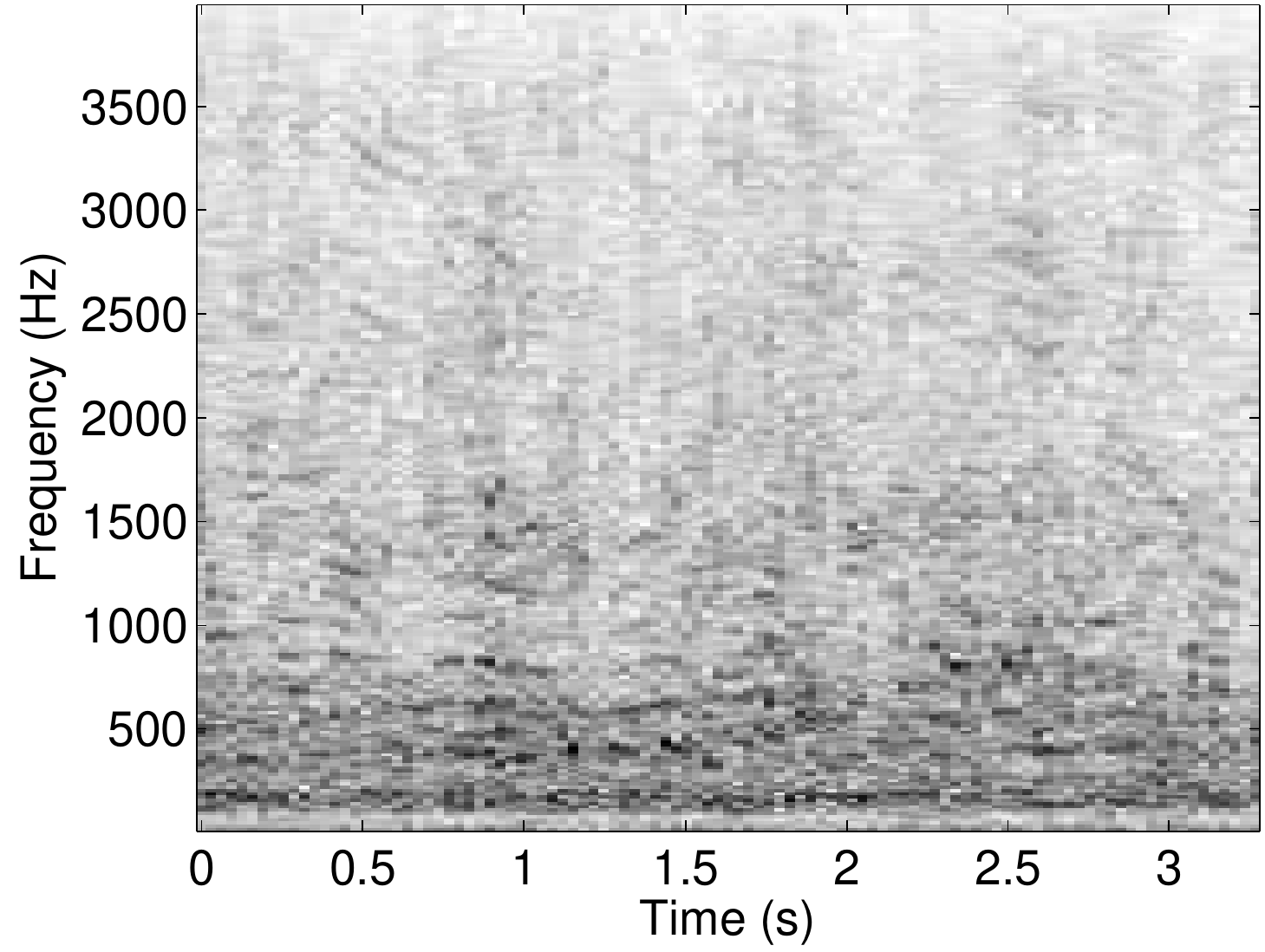}
	}
	\subfigure[Recovered noise]{
		\includegraphics[height=0.97in]{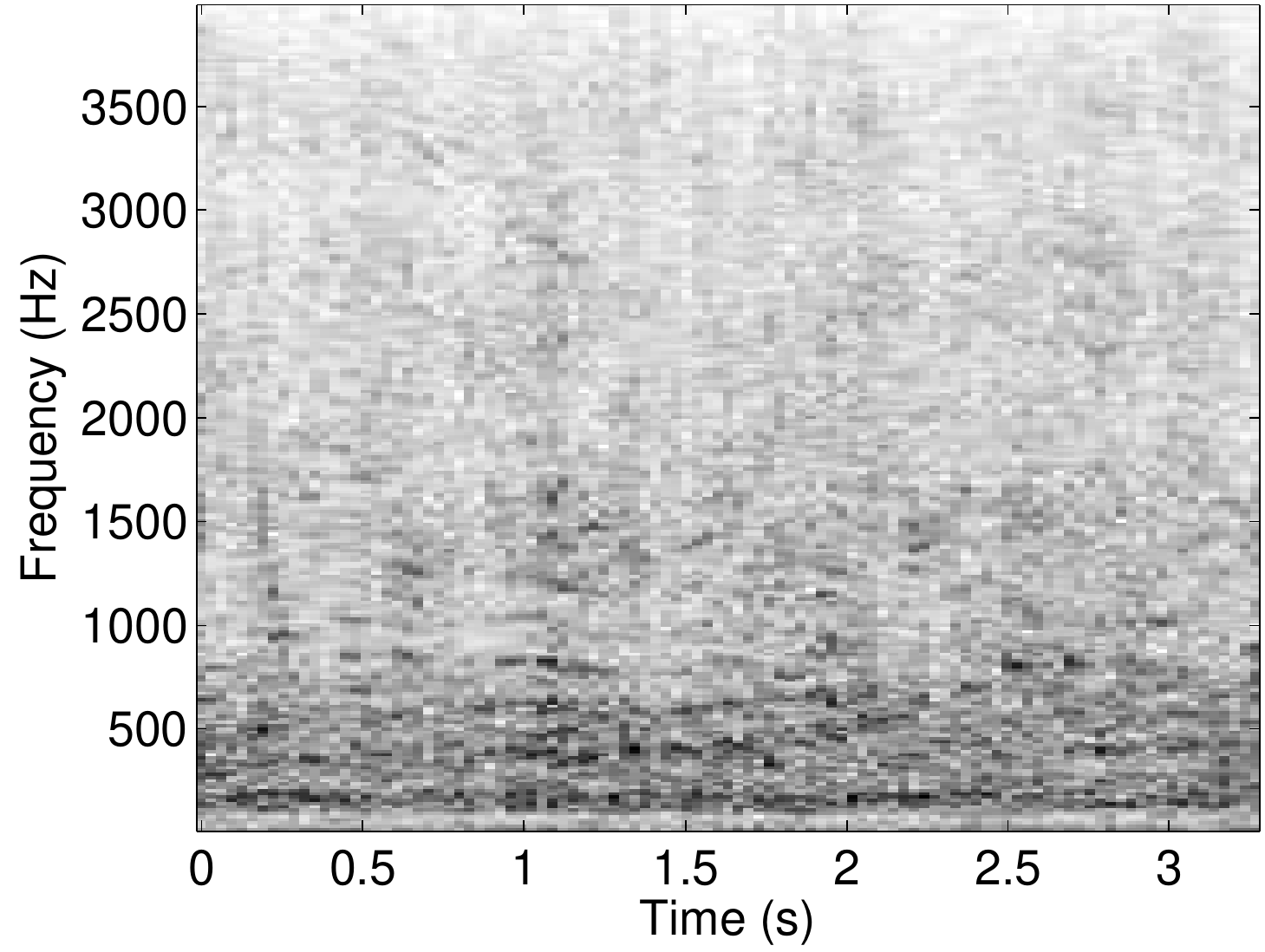}
	}
	\vspace{-4.mm}	
	\caption{A speech denoising example using the TIMIT dataset. {(a) The mixture (speech and babble noise) magnitude spectrogram for a test clip in TIMIT; (b) the ground truth spectrogram for the speech; (c) the separated speech spectrogram from our proposed model (DNN); (d) the ground truth spectrogram for the babble noise; (e) the separated babble noise spectrogram from our proposed model (DNN).}} 
	\label{fig:timit_separation_example}
	\vspace{-2mm}	
\end{figure*}

In Eq. \eqref{eq:mse}, we measure the difference between the predicted and the actual targets.
When targets have similar spectra, it is possible for the DNN to
minimize Eq. \eqref{eq:mse} by being too conservative: when a feature could be
attributed to either source 1 or source 2, the neural network
attributes it to both. The conservative strategy is effective in
training, but leads to reduced signal-to-interference ratio (SIR) in
testing, as the network allows ambiguous spectral features to \textit{bleed}
through partially from one source to the other. 
We address this issue by proposing a discriminative network training criterion for reducing the interference, possibly at the cost of increased artifacts.
Suppose that we define 
\begin{equation}
J_{\mathrm{DIS}} = -(1-\gamma)\ln p_{12}(\mathbf{y})-
\gamma D_\mathrm{KL}(p_{12}\Vert p_{21})
\label{eq:JDIS}
\end{equation}
where $0\le\gamma\le 1$ is a regularization constant. 
$p_{12}(\mathbf{y})$ is the likelihood of the training data under the
assumption that the neural net computes the MSE estimate of each
feature vector (i.e., its conditional expected value given knowledge
of the mixture), and that all residual noise is Gaussian with unit
covariance, thus
\begin{equation}
\ln p_{12}(\mathbf{y}) =
-\frac{1}{2}\sum_{t=1}^T \left(
\Vert\mathbf{y}_{\mathbf{1}_t}-\tilde{\mathbf{y}}_{\mathbf{1}_t}\Vert^2+
\Vert\mathbf{y}_{\mathbf{2}_t}-\tilde{\mathbf{y}}_{\mathbf{2}_t}\Vert^2
\right)
\label{eq:loglikelihood}
\end{equation}
The discriminative term, $D_\mathrm{KL}(p_{12}\Vert p_{21})$, is a point estimate of 
the KL divergence between the likelihood model $p_{12}(\mathbf{y})$ and
the model $p_{21}(\mathbf{y})$, where the latter is computed by swapping
affiliation of spectra to sources, thus
\begin{align*}
D_\mathrm{KL}(p_{12}\Vert p_{21}) &=&\frac{1}{2}\sum_{t=1}^T \big(
\Vert\mathbf{y}_{\mathbf{1}_t}-\tilde{\mathbf{y}}_{\mathbf{2}_t}\Vert^2+
\Vert\mathbf{y}_{\mathbf{2}_t}-\tilde{\mathbf{y}}_{\mathbf{1}_t}\Vert^2-\\
&&\Vert\mathbf{y}_{\mathbf{1}_t}-\tilde{\mathbf{y}}_{\mathbf{1}_t}\Vert^2-
\Vert\mathbf{y}_{\mathbf{2}_t}-\tilde{\mathbf{y}}_{\mathbf{2}_t}\Vert^2\big)
\numberthis
\label{eq:KLD}
\end{align*}

Combining Eqs.~\eqref{eq:JDIS}--\eqref{eq:KLD} gives a discriminative 
criterion with a simple and useful form:
\begin{align*}
J_{\mathrm{DIS}}&=&
\frac{1}{2}\sum_{t=1}^T \big(\Vert\mathbf{y}_{\mathbf{1}_t}-\tilde{\mathbf{y}}_{\mathbf{1}_t}\Vert^2+
\Vert\mathbf{y}_{\mathbf{2}_t}-\tilde{\mathbf{y}}_{\mathbf{2}_t}\Vert^2-\\
&&\gamma\Vert\mathbf{y}_{\mathbf{1}_t}-\tilde{\mathbf{y}}_{\mathbf{2}_t}\Vert^2-
\gamma\Vert\mathbf{y}_{\mathbf{2}_t}-\tilde{\mathbf{y}}_{\mathbf{1}_t}\Vert^2\big)
\numberthis
\label{eq:discrim_mse}
\end{align*}
Although Eq. \eqref{eq:mse} directly optimizes the reconstruction objective, adding the extra term $-\gamma\Vert\mathbf{y}_{\mathbf{1}_t}-\tilde{\mathbf{y}}_{\mathbf{2}_t}\Vert^2-
\gamma\Vert\mathbf{y}_{\mathbf{2}_t}-\tilde{\mathbf{y}}_{\mathbf{1}_t}\Vert^2$ in Eq. \eqref{eq:discrim_mse} further penalizes the interference from the other source, and can be viewed as a regularizer of Eq. \eqref{eq:mse} during the training. From our experimental results, we generally achieve higher source to interference ratio while maintaining similar or higher source to distortion ratio and source to artifacts ratio.

\begin{figure*}[ht!]
	\hspace{-8mm}
	\includegraphics[width=7.3in]{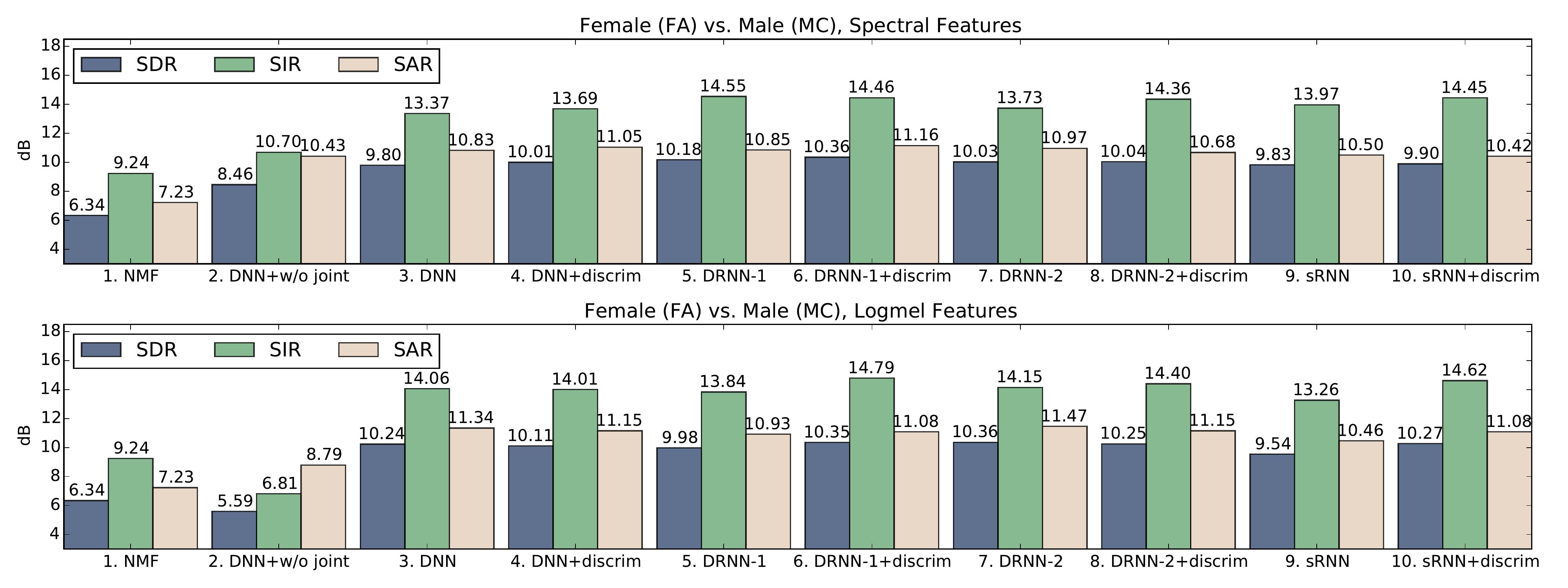}
	\caption{{TSP speech separation results (Female vs. Male), where ``w/o joint'' indicates the network is not trained with the masking layer, and ``discrim'' indicates the training with the discriminative objective. Note that the NMF model uses spectral features.}}
	\label{fig:TSP_d0_separation}
\end{figure*}
\begin{figure*}[th!]
	\hspace{-8mm}
	\includegraphics[width=7.3in]{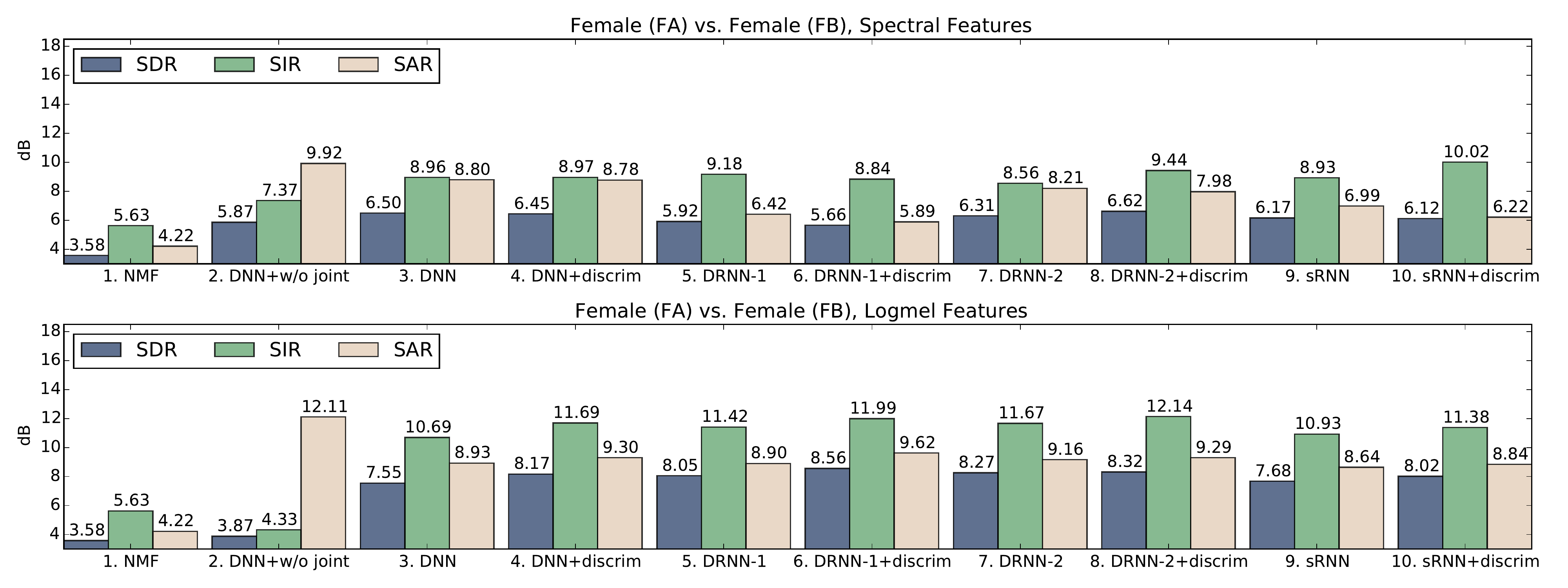}
	\caption{{TSP speech separation results (Female vs. Female), where ``w/o joint'' indicates the network is not trained with the masking layer, and ``discrim'' indicates the training with the discriminative objective. Note that the NMF model uses spectral features.}}
	\label{fig:TSP_d1_separation}
	
\end{figure*}
\section{Experiments}
\label{sec:exp} 
In this section, we evaluate the proposed models on three monaural source separation tasks: speech separation, singing voice separation, and speech denoising. 
We quantitatively evaluate the source separation performance using three metrics: Source to Interference Ratio (SIR), Source to Artifacts Ratio (SAR), and Source to Distortion Ratio (SDR), according to the BSS-EVAL metrics \cite{BSSEVAL}.
SDR is the ratio of the power of the input signal to the power of the difference between input and reconstructed signals. SDR is therefore exactly the same as the classical measure ``signal-to-noise ratio'' (SNR), and SDR reflects the overall separation performance. 
In addition to SDR, SIR reports errors caused by failures to fully remove the interfering signal, and SAR reports errors caused by extraneous artifacts introduced during the source separation procedure. 
In the past decade, the source separation community has been seeking more precise information about source reconstruction performance;
in particular, recent papers \cite{Liu_denoising_interspeech14, Bruna_NMF_2015} and competitions (e.g., Signal Separation Evaluation Campaign (SiSEC), Music Information Retrieval Evaluation (MIREX)) now separately report SDR, SIR, and SAR for objectively comparing different approaches.
Note that these measures are defined so that distortion = interference + artifacts. 
For the speech denoising task, we additionally compute the short-time objective intelligibility measure (STOI) which is a quantitative estimate of the intelligibility of the denoised speech \cite{Taal_STOI_ASLP11}.
Higher values of SDR, SAR, SIR, and STOI represent higher separation quality.

We use the abbreviations DRNN-$k$ and sRNN to denote the DRNN with the recurrent connection at the $k$-th hidden layer, or at all hidden layers, respectively. Examples are shown in Figure \ref{fig:DRNN_architectures}. We select the architecture and hyperparameters (the $\gamma$ parameter in Eq. \eqref{eq:discrim_mse}, the mini-batch size, L-BFGS iterations, and the circular shift size of the training data) based on the development set performance.

We optimize our models by back-propagating the gradients with respect to the training objective in Eq. \eqref{eq:discrim_mse}. We use the limited-memory Broyden-Fletcher-Goldfarb-Shanno (L-BFGS) algorithm \cite{Byrd1995LBFGS} to train the models from random initialization.
Examples of the separation results are shown in Figures \ref{fig:TSP_separation_example}, \ref{fig:mir_separation_example}, and \ref{fig:timit_separation_example}.
The sound examples and source codes of this work are available online.\footnote{https://sites.google.com/site/deeplearningsourceseparation/}

\begin{figure*}[th!]
	\hspace{-8mm}
	\includegraphics[width=7.3in]{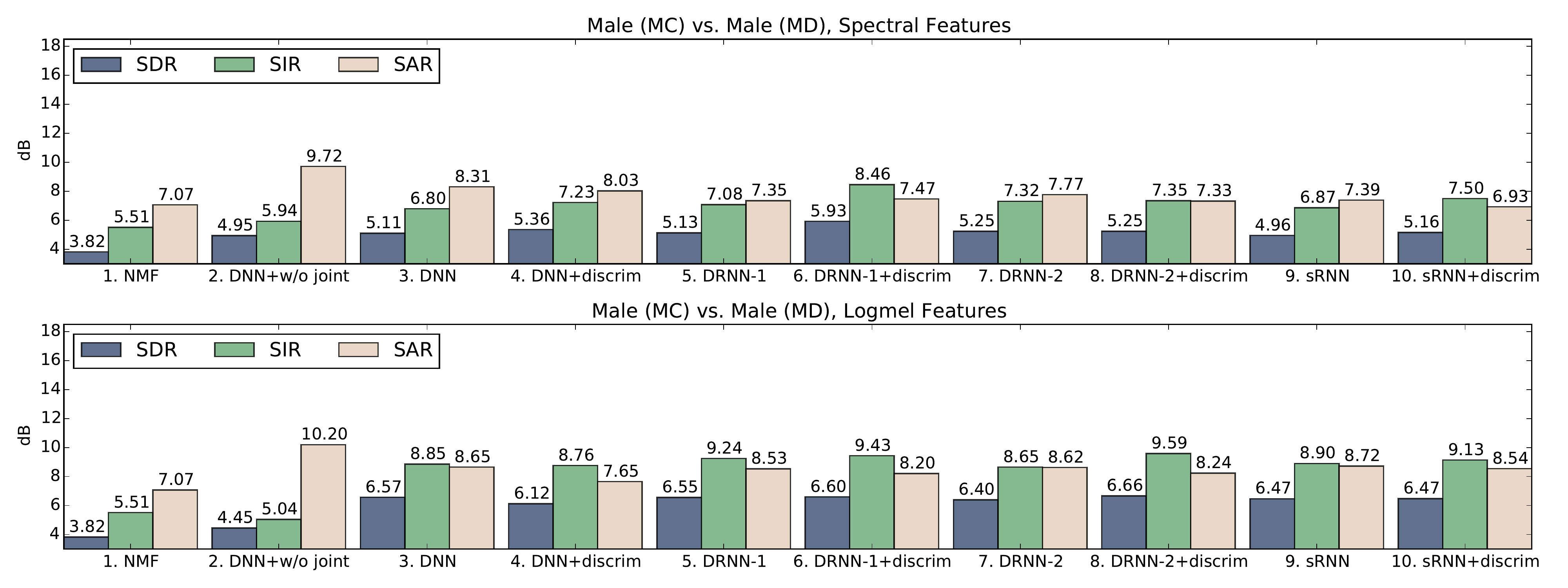}
	\caption{{TSP speech separation results (Male vs. Male), where ``w/o joint'' indicates the network is not trained with the masking layer, and ``discrim'' indicates the training with the discriminative objective. Note that the NMF model uses spectral features.}}
	\label{fig:TSP_d2_separation}
\end{figure*}

\subsection{Speech Separation Setting}
We evaluate the performance of the proposed approaches for a monaural speech separation task using the TSP corpus \cite{kabal2002tsp}. There are 1444 utterances, with average length 2.372 s, spoken by 24 speakers (half male and half female).
We choose four speakers, \texttt{FA} (female), \texttt{FB} (female), \texttt{MC} (male), and \texttt{MD} (male), from the TSP speech database. After concatenating together 60 sentences for each speaker, we use 80\% of the signals for training, 10\% for development, and 10\% for testing. 
The signals are downsampled to 16 kHz. 
The neural networks are trained on three different mixing cases: FA versus MC, FA versus FB, and MC versus MD. Since FA and FB are female speakers while MC and MD are male, the latter two cases are expected to be more difficult due to the similar frequency ranges from the same gender. After normalizing the signals to have 0 dB input SNR, the neural networks are trained to learn the mapping between an input mixture spectrum and the corresponding pair of clean spectra.

As for the NMF experiments, 10 to 100 speaker-specific basis vectors are trained from the training part of the signals. 
The optimal number of basis vectors is chosen based on the development set. We empirically found that using 20 basis vectors achieves the best performance on the development set in the three different mixing cases.
The NMF separation is done by fixing the known speakers' basis vectors during the test procedure and learning the speaker-specific activation matrices. 

We explore two different types of input features: spectral and log-mel filterbank features. The spectral representation is extracted using a 1024-point shot-time Fourier transform (STFT) with 50\% overlap. 
In the speech recognition literature \cite{li2012improving}, the log-mel filterbank is found to provide lower word-error-rate compared to mel-frequency cepstral coefficients (MFCC) and log FFT bins. The 40-dimensional log-mel representation and the first- and second-order derivative features are used in the experiments. 
For the neural network training, in order to increase the variety of training samples, we circularly shift (in the time domain) the signals of one speaker and mix them with utterances from the other speaker. 

\subsection{Speech Separation Results}

We use the standard NMF with the generalized KL-divergence metric as our baseline. We report the best NMF results among models with different basis vectors, as shown in the first column of Figures \ref{fig:TSP_d0_separation}, \ref{fig:TSP_d1_separation}, and \ref{fig:TSP_d2_separation}. Note that NMF uses spectral features, and hence the results in the second row (log-mel features) of each figure are the same as the first row (spectral features).

The speech separation results of the cases, FA versus MC, FA versus FB, and MC versus MD, are shown in Figures \ref{fig:TSP_d0_separation}, \ref{fig:TSP_d1_separation}, and \ref{fig:TSP_d2_separation}, respectively. 
We train models with two hidden layers of 300 hidden units using features with a context window size of one frame (one frame within a window), where the architecture and the hyperparameters are chosen based on the development set performance. 
We report the results of single frame spectra and log-mel features in the top and bottom rows of Figures \ref{fig:TSP_d0_separation}, \ref{fig:TSP_d1_separation}, and \ref{fig:TSP_d2_separation}, respectively. To further understand the strength of the models, we compare the experimental results in several aspects. 
In the second and third columns of Figures \ref{fig:TSP_d0_separation}, \ref{fig:TSP_d1_separation}, and \ref{fig:TSP_d2_separation}, we examine the effect of joint optimization of the masking layer and the DNN. Jointly optimizing the masking layer significantly outperforms the cases where the masking layer is applied separately (the second column). 
In the FA vs. FB case, DNN without joint optimization of the masking layer achieves high SAR, but results in low SDR and SIR. 
In the top and bottom rows of Figures \ref{fig:TSP_d0_separation}, \ref{fig:TSP_d1_separation}, and \ref{fig:TSP_d2_separation}, we compare the results between spectral features and log-mel features. 
In the joint optimization case, (columns 3--10), log-mel features achieve higher SDRs, SIRs, and SARs compared to spectral features. On the other hand, spectral features achieve higher SDRs and SIRs in the case where DNN is not jointly trained with a masking layer, as shown in the second column of Figures \ref{fig:TSP_d0_separation}, \ref{fig:TSP_d1_separation}, and \ref{fig:TSP_d2_separation}. In the FA vs. FB and MC vs. MD cases, the log-mel features outperform spectral features greatly. 

Between columns 3, 5, 7, and 9, and columns 4, 6, 8, and 10 of Figures \ref{fig:TSP_d0_separation}, \ref{fig:TSP_d1_separation}, and \ref{fig:TSP_d2_separation}, we make comparisons between various network architectures, including DNN, DRNN-1, DRNN-2, and sRNN. In many cases, recurrent neural network models (DRNN-1, DRNN-2, or sRNN) outperform DNN. 
Between columns 3 and 4, columns 5 and 6, columns 7 and 8, and columns 9 and 10 of Figures \ref{fig:TSP_d0_separation}, \ref{fig:TSP_d1_separation}, and \ref{fig:TSP_d2_separation}, we compare the effectiveness of using the discriminative training criterion, i.e., $\gamma>0$ in Eq. \eqref{eq:discrim_mse}. In most cases, SIRs are improved. The results match our expectation when we design the objective function. However, it also leads to some artifacts which result in slightly lower SARs in some cases.
Empirically, the value $\gamma$ is in the range of 0.01--0.1 in order to achieve SIR improvements and maintain reasonable SAR and SDR.

\begin{figure}[t!]
  	\hspace{-3mm}
  	 \includegraphics[width=3.5in]{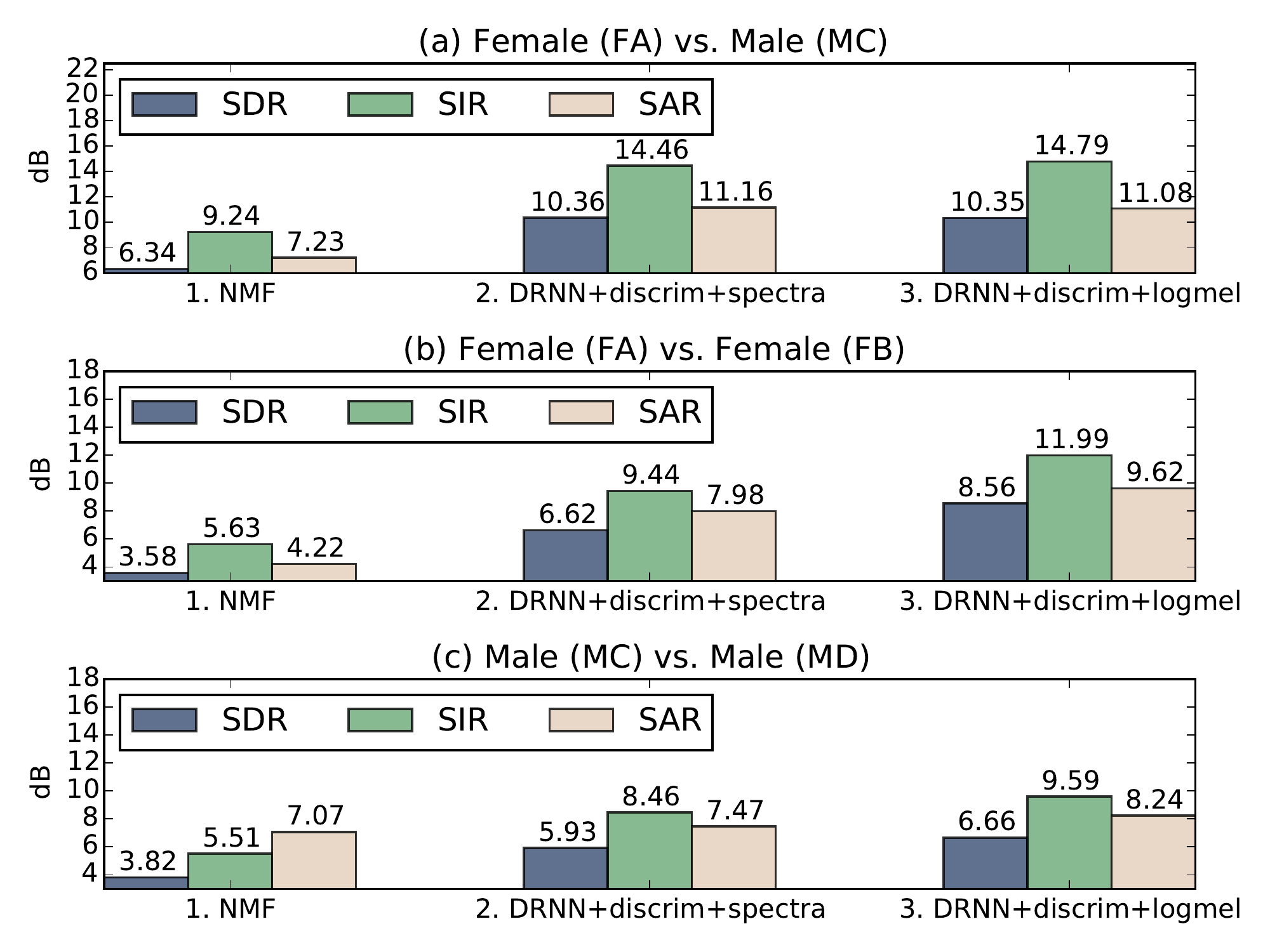}
  	\vspace{-4mm}
  	\caption{{TSP speech separation result summary. We compare the results under three settings, (a) Female vs. Male, (b) Female vs. Female, and (c) Male vs. Male, using the NMF model, the best DRNN+discrim architecture with spectra features, and the best DRNN+discrim architecture with log-mel features.}}
  	\label{fig:TSP_all_separation}
  \end{figure}

Finally, we compare the NMF results with our proposed models with the best architecture using spectral and log-mel features, as shown in Figure \ref{fig:TSP_all_separation}. NMF models learn activation matrices from different speakers and hence perform poorly in the same sex speech separation cases, FA vs. FB and MC vs. MD. 
Our proposed models greatly outperform NMF models for all three cases. Especially for the FA vs. FB case, our proposed model achieves around 5 dB SDR gain compared to the NMF model while maintaining higher SIR and SAR.

  \vspace{-1mm}
\subsection{Singing Voice Separation Setting}
We apply our models to a singing voice separation task, where one source is the singing voice and the other source is the background music. The goal is to separate singing voice from music recordings.

We evaluate our proposed system using the MIR-1K dataset \cite{Chao-Ling_Jang_2010}.\footnote{https://sites.google.com/site/unvoicedsoundseparation/mir-1k} A thousand song clips are encoded at a sampling rate of 16 KHz, with a duration from 4 to 13 seconds. The clips were extracted from 110 Chinese karaoke songs performed by both male and female amateurs. There are manual annotations of the pitch contours, lyrics, indices and types for unvoiced frames, and the indices of the vocal and non-vocal frames; none of the annotations were used in our experiments. Each clip contains the singing voice and the background music in different channels.

Following the evaluation framework in \cite{Pablo_lowrank_2012, Yang_2013_LowReprOfBoth}, we use 175 clips sung by one male and one female singer (``abjones'' and ``amy'') as the training and development set.\footnote{Four clips, abjones\_5\_08, abjones\_5\_09, amy\_9\_08, amy\_9\_09, are used as the development set for adjusting the hyperparameters.} The remaining 825 clips of 17 singers are used for testing. For each clip, we mixed the singing voice and the background music with equal energy, i.e., 0 dB SNR. 

To quantitatively evaluate the source separation results, 
we report the overall performance via Global NSDR (GNSDR), Global SIR (GSIR), and Global SAR (GSAR), which are the weighted means of the NSDRs, SIRs, SARs, respectively, over all test clips weighted by their length. Normalized SDR (NSDR) \cite{Ozerov_NSDR_2007} is defined as:
\begin{equation}
\label{ }
\text{NSDR}(\hat{\mathbf{v}},\mathbf{v},\mathbf{x})=\text{SDR}(\hat{\mathbf{v}},\mathbf{v})-\text{SDR}(\mathbf{x},\mathbf{v})
\end{equation}
where $\hat{\mathbf{v}}$ is the estimated singing voice, $\mathbf{v}$ is the original clean singing voice, and $\mathbf{x}$ is the mixture. NSDR is for estimating the improvement of the SDR between the preprocessed mixture $\mathbf{x}$ and the separated singing voice $\hat{\mathbf{v}}$.

For the neural network training, in order to increase the variety of training samples, we circularly shift (in the time domain) the signals of the singing voice and mix them with the background music.
In the experiments, we use magnitude spectra as input features to the neural network. The spectral representation is extracted using a 1024-point STFT with 50\% overlap. Empirically, we found that using log-mel filterbank features or log power spectrum provide worse performance than using magnitude spectra in the singing voice separation task.
\vspace{-1mm}
\subsection{Singing Voice Separation Results}
In this section, we compare various deep learning models from several aspects, including the effect of different output formats, the effect of different deep recurrent neural network structures, and the effect of discriminative training.

For simplicity, unless mentioned explicitly, we report the results using three hidden layers of 1000 hidden units deep neural networks with the mean squared error criterion, joint optimization of the masking layer, and 10 K samples as the circular shift step size using features with a context window size of three frames (three frames within a window).





Table \ref{tab:output_format_results} presents the results with different output layer formats. We compare using single source as a target (row 1) and using two sources as targets in the output layer (row 2 and row 3). We observe that modeling two sources simultaneously provides higher performance in GNSDR, GSIR, and GSAR. Comparing row 2 and row 3 in Table \ref{tab:output_format_results}, we observe that jointly optimizing the masking layer and the DRNN further improves the results.

\begin{table}[t!]
\begin{center}
\setlength
{\tabcolsep}{0.5em}
  \caption{MIR-1K separation result comparison using deep neural networks with single source as a target and using two sources as targets (with and without joint optimization of the masking layers and the DNNs).}
  \label{tab:output_format_results}
  \vspace{-1mm}
  \begin{tabular}{|c|c|c|c|}
  \hline
  Model (num. of output  &  \multirow{2}{*}{GNSDR} &  \multirow{2}{*}{GSIR} & \multirow{2}{*}{GSAR} \\
  sources, joint optimization) &&&\\\hline
  DNN (1, no) & 5.64 & 8.87 & 9.73 \\\hline
  DNN (2, no) & 6.44 & 9.08 & 11.26\\\hline
  DNN (2, yes) & 6.93 & 10.99 & 10.15\\\hline
\end{tabular}

\end{center}
\end{table}


Table \ref{tab:DRNN_discrim_results} presents the results of different deep recurrent neural network architectures (DNN, DRNN with different recurrent connections, and sRNN) with and without discriminative training. We can observe that discriminative training further improves GSIR while maintaining similar GNSDR and GSAR.

		\begin{table}[t]
		\begin{center}
		\setlength
		{\tabcolsep}{0.5em}
		\caption{MIR-1K separation result comparison for the effect of discriminative training using different architectures. ``discrim'' denotes the models with discriminative training.}
		\vspace{-1mm}
		\label{tab:DRNN_discrim_results}
		  \begin{tabular}{|c|c|c|c|}
		  \hline
		  Model & GNSDR & GSIR& GSAR \\\hline   
		  DNN  & 6.93 & 10.99 & 10.15 \\\hline
		  DRNN-1 & 7.11 & 11.74 & 9.93\\\hline
		  DRNN-2 & 7.27 & 11.98 & 9.99\\\hline
		  DRNN-3 & 7.14 & 11.48 & 10.15\\\hline   
		  sRNN & 7.09 & 11.72 & 9.88\\\hline\hline
		  DNN + discrim & 	7.09 & 12.11 & 9.67 \\\hline
		  DRNN-1 + discrim & 7.21 & 12.76 & 9.56\\\hline
		  DRNN-2 + discrim & 7.45 & 13.08 & 9.68\\\hline
		  DRNN-3 + discrim & 7.09 & 11.69 & 10.00\\\hline   
		  sRNN + discrim & 7.15 & 12.79 & 9.39\\\hline

		\end{tabular}

		\end{center}
		\end{table}		

\begin{table}[ht!]
	\begin{center}
		\setlength
		{\tabcolsep}{0.5em}
		\caption{MIR-1K separation result comparison between our models and previous proposed approaches. ``discrim'' denotes the models with discriminative training.}
		\label{tab:overall_results}
		\begin{tabular}{|c|c|c|c|}
			\hline
			\multicolumn{4}{|c|}{Unsupervised}\\\hline
			Model & GNSDR & GSIR& GSAR \\\hline   
			RPCA  \cite{Huang_RPCA_Separation_ICASSP2012} & 3.15 & 4.43 & 11.09 \\\hline
			RPCAh \cite{Yang_sparse_lowrank_2012} & 3.25 & 4.52 & 11.10\\\hline
			RPCAh + FASST \cite{Yang_sparse_lowrank_2012} & 3.84 & 6.22 & 9.19\\\hline\hline
			
			\multicolumn{4}{|c|}{Supervised}\\\hline
			Model & GNSDR & GSIR& GSAR \\\hline   
			MLRR \cite{Yang_2013_LowReprOfBoth} & 3.85 & 5.63 & 10.70\\\hline
			RNMF \cite{Pablo_lowrank_2012} & 4.97 & 7.66 & 10.03\\\hline  
			{\bf DRNN-2} & {\bf 7.27} & {\bf 11.98} & {\bf 9.99}\\\hline
			{\bf DRNN-2 + discrim} & {\bf 7.45} & {\bf 13.08} & {\bf 9.68}\\\hline
		\end{tabular}
		
	\end{center}
	
\end{table}	
Finally, we compare our best results with other previous work under the same setting. Table \ref{tab:overall_results} shows the results with unsupervised and supervised settings. Our proposed models achieve 2.30--2.48 dB GNSDR gain, 4.32--5.42 dB GSIR gain with similar GSAR performance, compared with the RNMF model \cite{Pablo_lowrank_2012}.

\subsection{Speech Denoising Setting}
We apply the proposed framework to a speech denoising task, where one source is the clean speech and the other source is the noise. The goal of the task is to separate clean speech from noisy speech.
In the experiments, we use magnitude spectra as input features to the neural network. The spectral representation is extracted using a 1024-point STFT with 50\% overlap. Empirically, we found that log-mel filterbank features provide worse performance than magnitude spectra.
Unless mentioned explicitly, we use two hidden layers of 1000 hidden units deep neural networks with the mean squared error criterion, joint optimization of the masking layer, and 10 K samples as the circular shift step size, using features with a context window size of one frame (one frame within a window). The model is trained and tested on 0 dB mixtures, without input normalization. 

To understand the effect of degradation in the mismatch condition, we set up the experimental recipe as follows. We use a hundred utterances spanning ten different speakers from the TIMIT database. We also use a set of five noises: Airport, Train, Subway, Babble, and Drill.
We generate a number of noisy speech recordings by selecting random subsets of noises and overlaying them with speech signals. 
We also specify the signal to noise ratio when constructing the noisy mixtures. After we complete the generation of the noisy signals, we split them into a training set and a test set. 

\subsection{Speech Denoising Results}
\begin{figure}[t!]
 	\begin{flushleft} 
 		\hspace{-5.5mm}
 \includegraphics[width=3.68in]{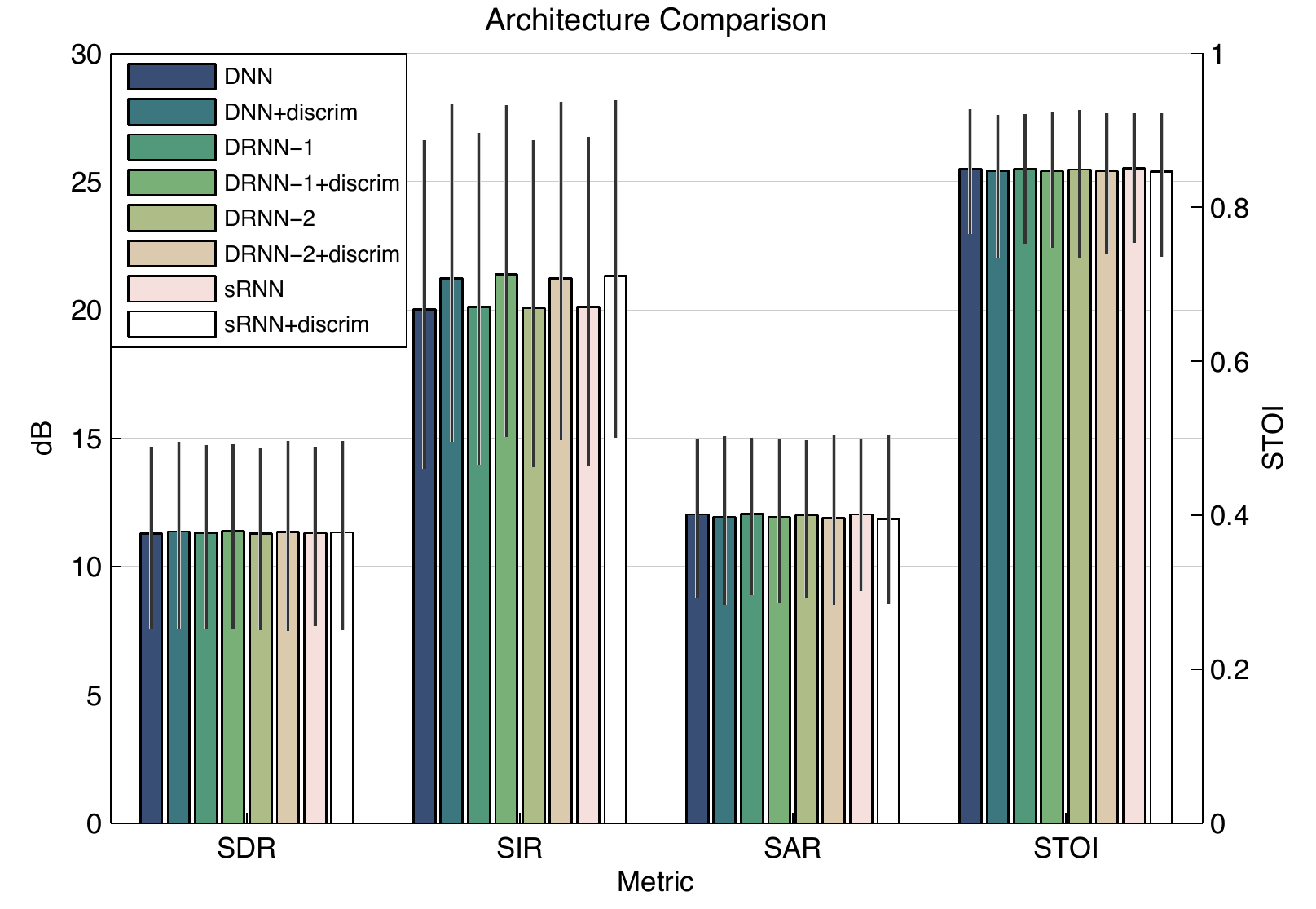}
 		\caption{Speech denoising architecture comparison, where ``+discrim" indicates the training with the discriminative objective, and the bars show average values and the vertical lines on the bars denote minimum and maximum observed values. Models are trained and tested on 0 dB SNR inputs. The average STOI score for unprocessed mixtures is 0.675.}
 		\label{fig:denoising_architecture}
 	\end{flushleft}
 \end{figure}

In the following experiments, we examine the effect of the proposed methods under various scenarios. 
We first evaluate various architectures using 0 dB SNR inputs, as shown in Figure \ref{fig:denoising_architecture}. 
We can observe that the recurrent neural network architectures (DRNN-1, DRNN-2, sRNN) achieve similar performance compared to the DNN model. Including the discriminative training objective improves SDR and SIR, but results in slightly degraded SAR and similar STOI values. 

\begin{figure}[t!]
	\centering
		\hspace{-1mm}
		 \includegraphics[width=3.7in]{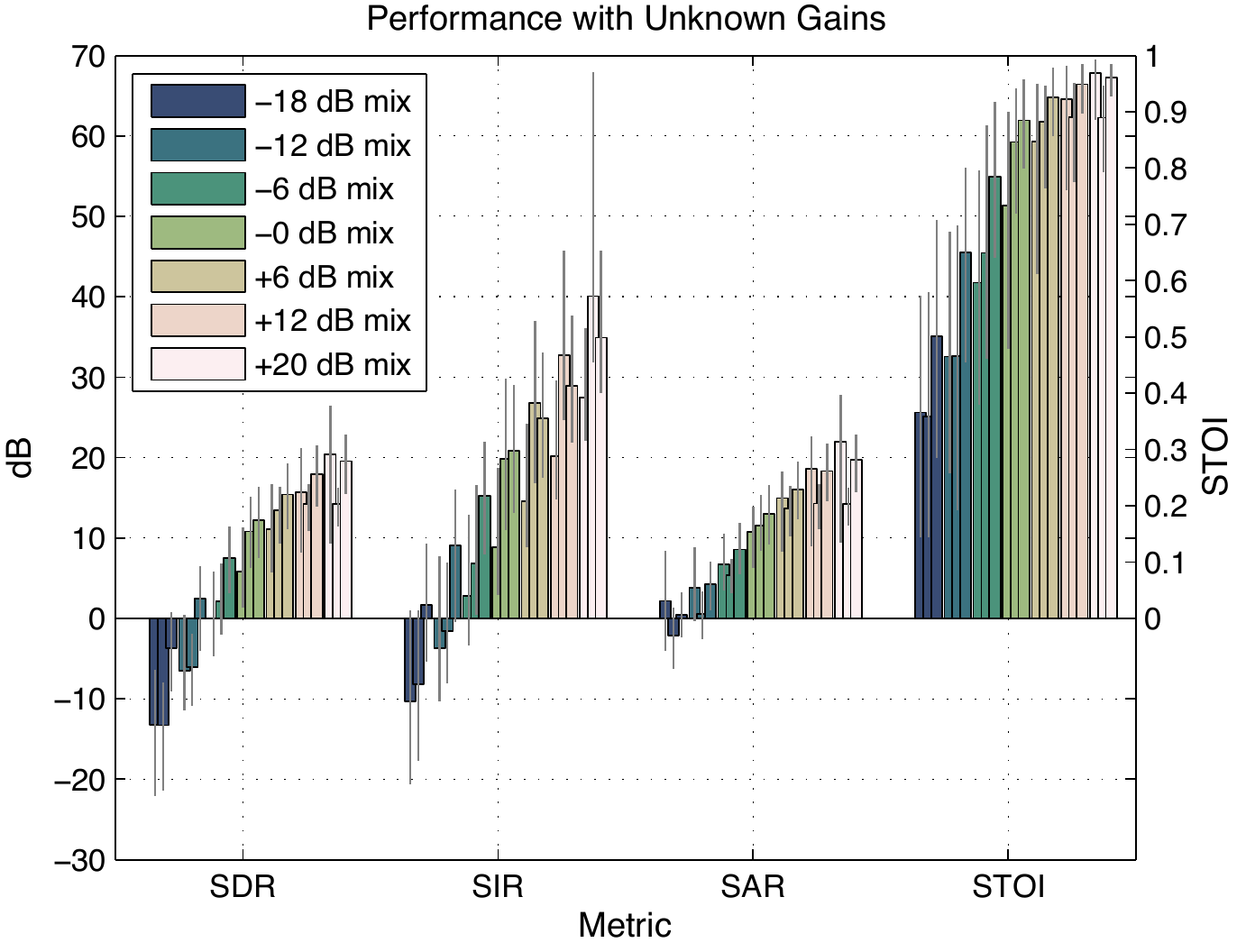}
		\caption{{Speech denoising using multiple SNR inputs and testing on a model that is trained on 0 dB SNR, where the bars show average values and the vertical lines on the bars denote minimum and maximum observed values. The left/back, middle, right/front bars in each pair show the results of NMF, DNN without joint optimization of the masking layer \cite{Liu_denoising_interspeech14}, and DNN with joint optimization of the masking layer, respectively. The average STOI scores for unprocessed mixtures at -18 dB, -12 dB, -6 dB, 0 dB, 6 dB, 12 dB, and 20 dB SNR are 0.370, 0.450, 0.563, 0.693, 0.815, 0.903, and 0.968, respectively.}}		 
		\label{fig:denoising_gain}
\end{figure}

\begin{figure*}[ht!]
	\centering
	\subfigure[Known speakers and noise]{
		\includegraphics[height=2.45in]{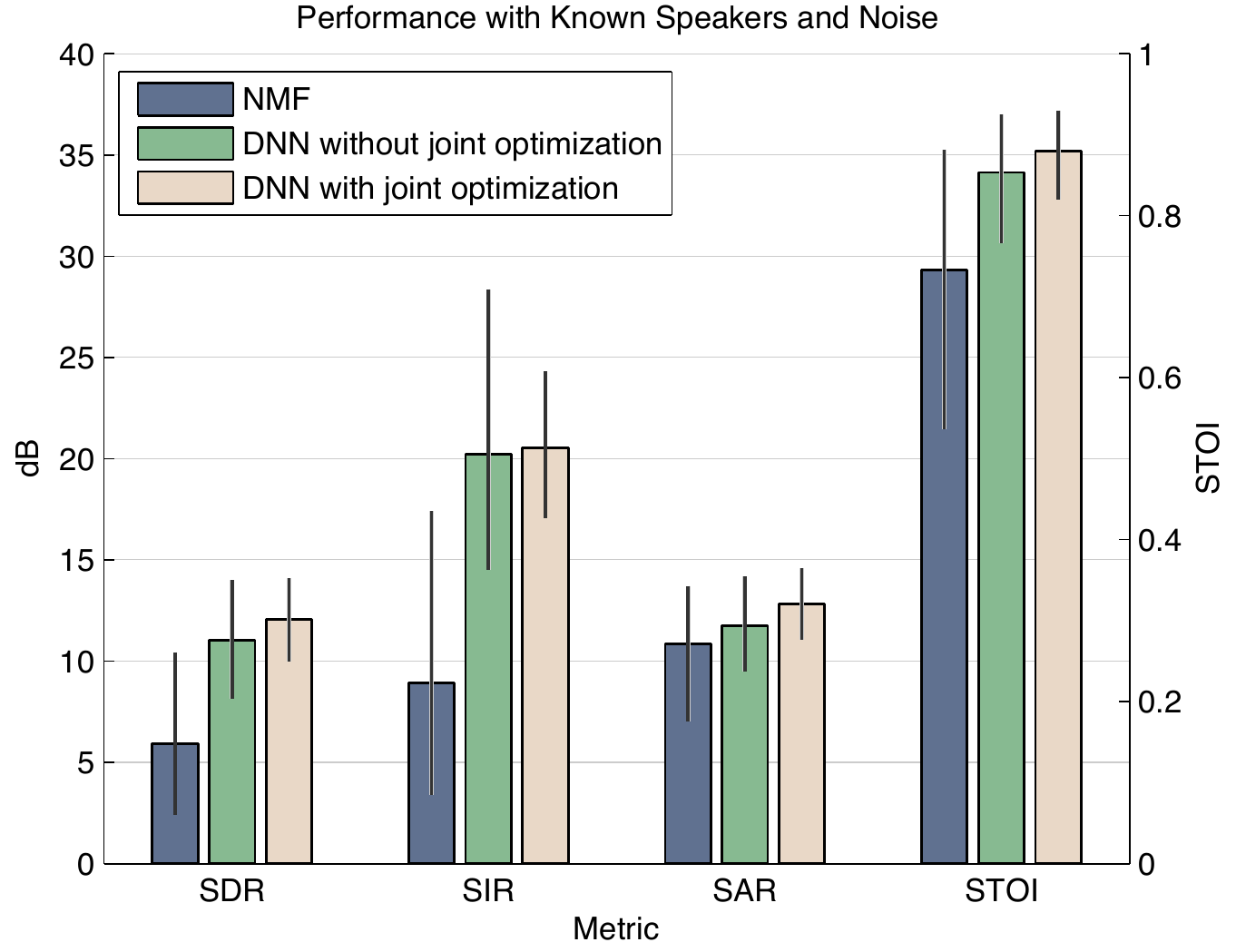}
	}
	\subfigure[Unknown speakers]{
		\includegraphics[height=2.45in]{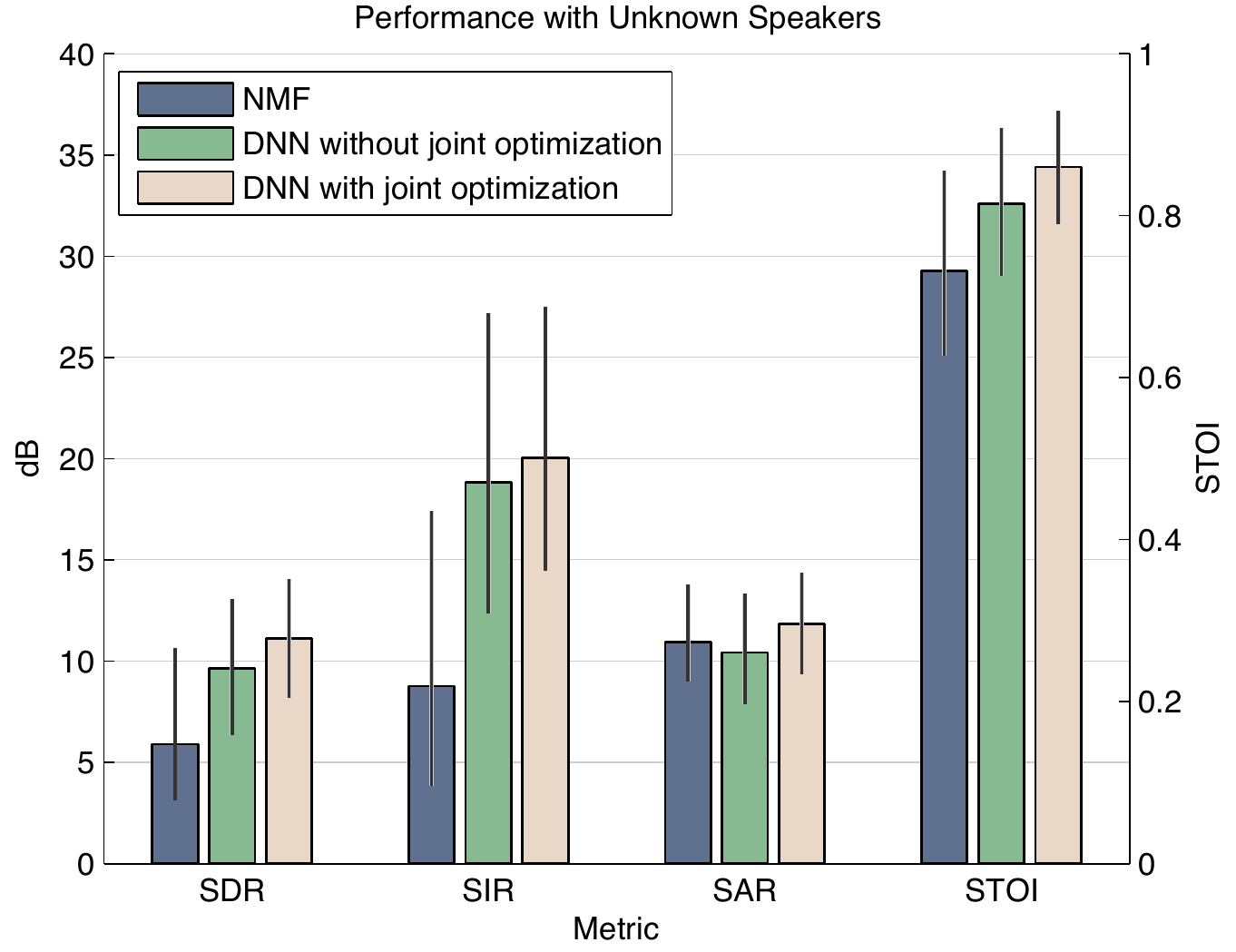}
	}
	\subfigure[Unknown noise]{
		\includegraphics[height=2.45in]{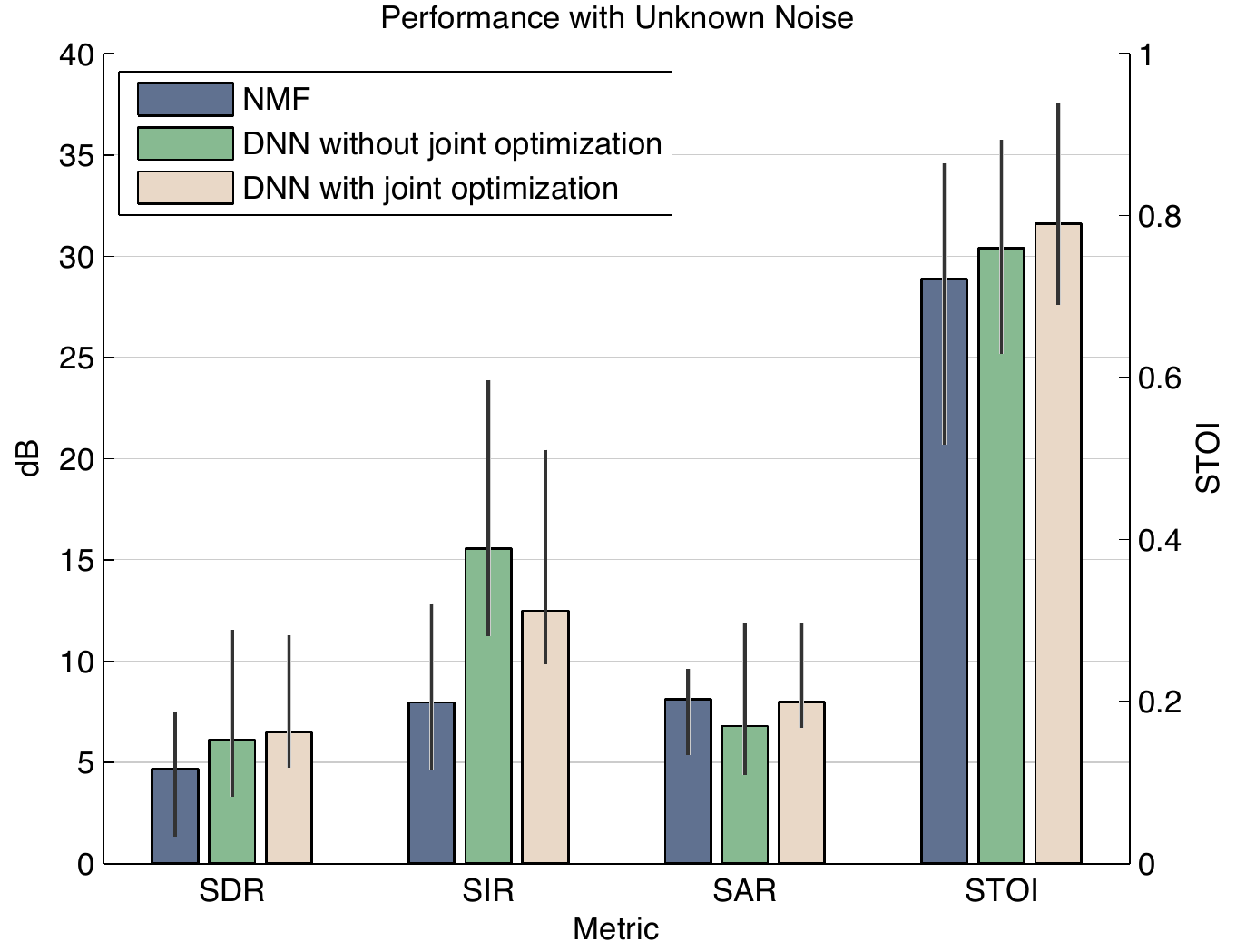}
	}
	\subfigure[Unknown speakers and noise]{
		\includegraphics[height=2.45in]{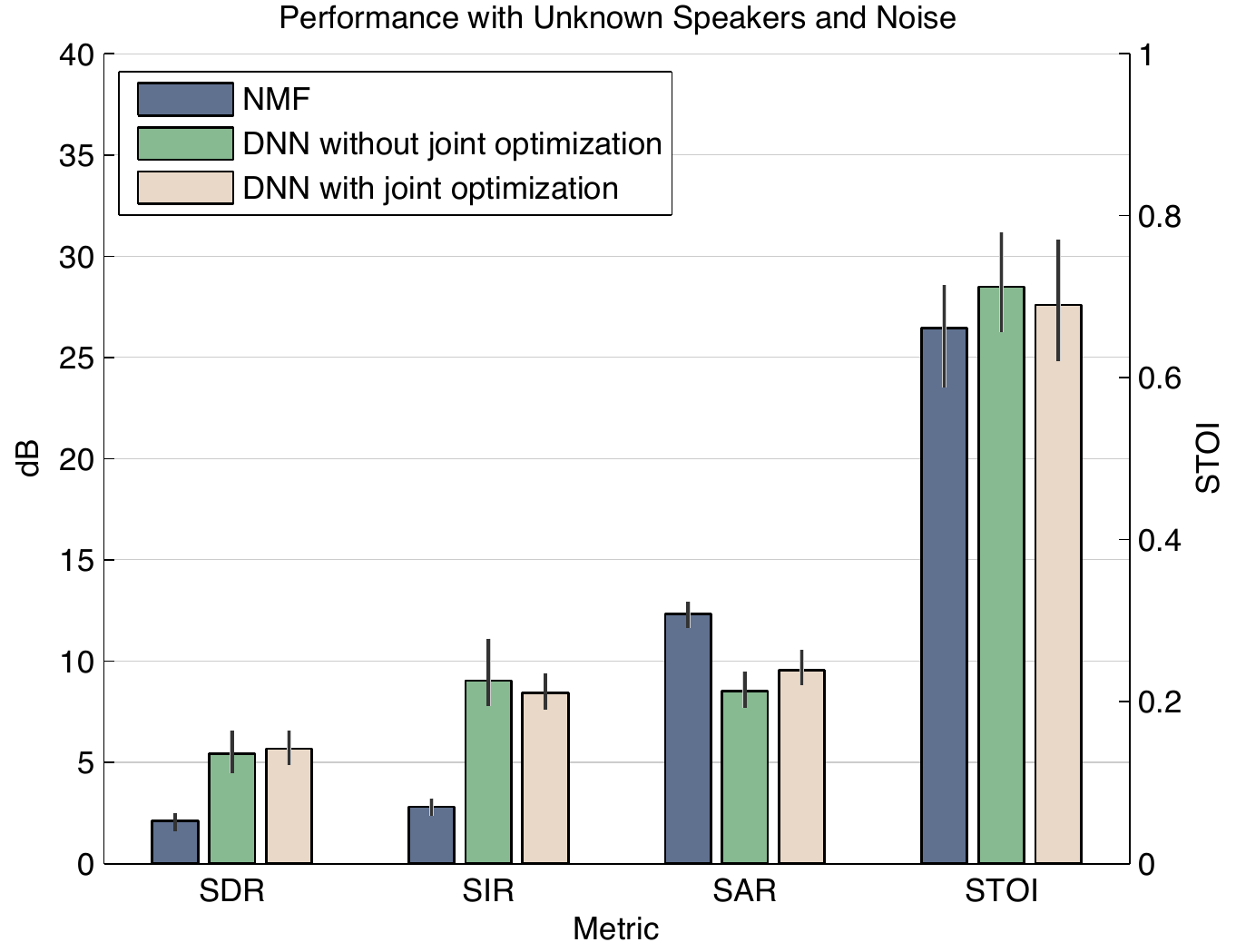}
	}\vspace{-2mm}
	\caption{{Speech denoising experimental results comparison between NMF, DNN without joint optimization of the masking layer \cite{Liu_denoising_interspeech14}, and DNN with joint optimization of the masking layer, given 0 dB SNR inputs, when used on data that is not represented in training. The bars show average values and the vertical lines on the bars denote minimum and maximum observed values. We show the separation results of (a) known speakers and noise, (b) unseen speakers, (c) unseen noise, and (d) unseen speakers and noise. The average STOI scores for unprocessed mixtures for cases (a), (b), (c), and (d) are 0.698, 0.686, 0.705, and 0.628, respectively.}} 
	\label{fig:denoising_unseen_data}
	\vspace{-1mm}	
\end{figure*}

To further evaluate the robustness of the model, we examine our model under a variety of situations in which it is presented with unseen data, such as unseen SNRs, speakers, and noise types. These tests provide a way of understanding the performance of the proposed approach under mismatched conditions. 
In Figure \ref{fig:denoising_gain}, we show the robustness of this model under various SNRs. The model is trained on 0 dB SNR mixtures and it is evaluated on mixtures ranging from 20 dB SNR to -18 dB SNR.
We compare the results between NMF, DNN without joint optimization of the masking layer, and DNN with joint optimization of the masking layer. In most cases, DNN with joint optimization achieves the best results, especially under low SNR inputs. For the 20 dB SNR case, NMF achieves the best performance. DNN without joint optimization achieves the highest SIR given high SNR inputs, though SDR, SAR, and STOI are lower than the DNN with joint optimization. 
Note that in our approach, joint optimization of the time-frequency masks and DNNs can be viewed as a way to directly incorporate the FFT-MASK targets \cite{Wang_Wang_target_2014} into the DNNs for both speech and noise, where authors in \cite{Wang_Wang_target_2014} found FFT-MASK has achieved better performance compared to other targets in speech denoising tasks. 

Next, we evaluate the models under three different cases: (1) the testing noise is unseen in training, (2) the testing speaker is unseen in training, and (3) both the testing noise and testing speaker are unseen in training stage.
For the unseen noise case, we train the model on mixtures with Babble, Airport, Train and Subway noises, and evaluate it on mixtures that include a Drill noise (which is significantly different from the training noises in both spectral and temporal structure).
For the unknown speaker case, we hold out some of the speakers from the training data. For the case where both the noise and speaker are unseen, we use the combination of the above. 

We compare our proposed approach with the NMF model and DNN without joint optimization of the masking layer \cite{Liu_denoising_interspeech14}. 
The models are trained and tested on 0 dB SNR inputs, and these experimental results are shown in Figure \ref{fig:denoising_unseen_data}. 
For the unknown speaker case, as shown in Figure \ref{fig:denoising_unseen_data} (b), we observe that there is only a mild degradation in performance for all models compared to the case where the speakers are known in Figure \ref{fig:denoising_unseen_data} (a). The results suggest that the approaches can be easily used in speaker variant situations. 
In Figure \ref{fig:denoising_unseen_data} (c), with the unseen noise, we observe a larger degradation in results, which is expected due to the drastically different nature of the noise type. 
For the case where both the noise and speakers are unknown, as shown in Figure \ref{fig:denoising_unseen_data} (d), all three models achieve the worst performance compare to the other cases. 
Overall, the proposed approach generalizes well across speakers and achieves higher source separation performance, especially in SDRs, compared to the baseline models under various conditions.

\subsection{Discussion}
Throughout the experiments in speech separation, singing voice separation, and speech denoising tasks, we have seen significant improvement over the baseline models under various settings, by the use of joint optimization of time-frequency masks with deep recurrent neural networks and the discriminative training objective.
By jointly optimizing time-frequency masks with deep recurrent neural networks, the proposed end-to-end system outperforms baseline models (such as NMF, DNN models without joint optimization) in matched and mismatched conditions.
Given audio signals are time series in nature, we explore various recurrent neural network architectures to capture temporal information and further enhance performance. Though 
there are extra memory and computational costs 
compared to feed-forward neural networks, DRNNs achieve extra gains, especially in the speech separation (0.5 dB SDR gain) and singing voice separation (0.34 dB GNSDR gain) tasks. Similar observations can be found in related work using LSTM models \cite{Weninger_LSTMseparation_ICASSP14, Weninger_LSTMdiscriminatively_14}, where the authors observe significant improvements using recurrent neural networks compared with DNN models.
Our proposed discriminative objective can be viewed as a regularization term towards the original mean-squared error objective. By enforcing the similarity between targets and predictions of the same source and dissimilarity between targets and predictions of competing sources, we observe that interference is further reduced while maintaining similar or higher SDRs and SARs.

\section{Conclusion and Future work} 
\label{sec:conclusion}
In this paper, we explore various deep learning architectures, including deep neural networks and deep recurrent neural networks for monaural source separation problems. We enhance the performance by jointly optimizing a soft time-frequency mask layer with the networks in an end-to-end fashion and exploring a discriminative training criterion. 
We evaluate our proposed method for speech separation, singing voice separation, and speech denoising tasks. Overall, our proposed models achieve 2.30--4.98 dB SDR gain compared to the NMF baseline, while maintaining higher SIRs and SARs in the TSP speech separation task.
In the MIR-1K singing voice separation task, our proposed models achieve 2.30--2.48 dB GNSDR gain and 4.32--5.42 dB GSIR gain, compared to the previously proposed methods, while maintaining similar GSARs.
Moreover, our proposed method also outperforms NMF and DNN baselines in various mismatch conditions in the TIMIT speech denoising task.
To further improve the performance, one direction is to further explore using LSTMs to model longer temporal information \cite{HochreiterLSTM_Neural97}, which has shown great performance compared to conventional recurrent neural networks as LSTM has properties of avoiding vanishing gradient properties. In addition, our proposed models can also be applied to many other applications such as robust ASR.

\appendices

\section*{Acknowledgment}
The authors would like to thank the anonymous reviewers for their valuable comments and suggestions. 

\ifCLASSOPTIONcaptionsoff
  \newpage
\fi

\bibliographystyle{IEEEtran}
\bibliography{DRNN_source_separation}

\vspace{-8mm}
\begin{IEEEbiography}[{\includegraphics[width=1in,height=1.25in,clip,keepaspectratio]{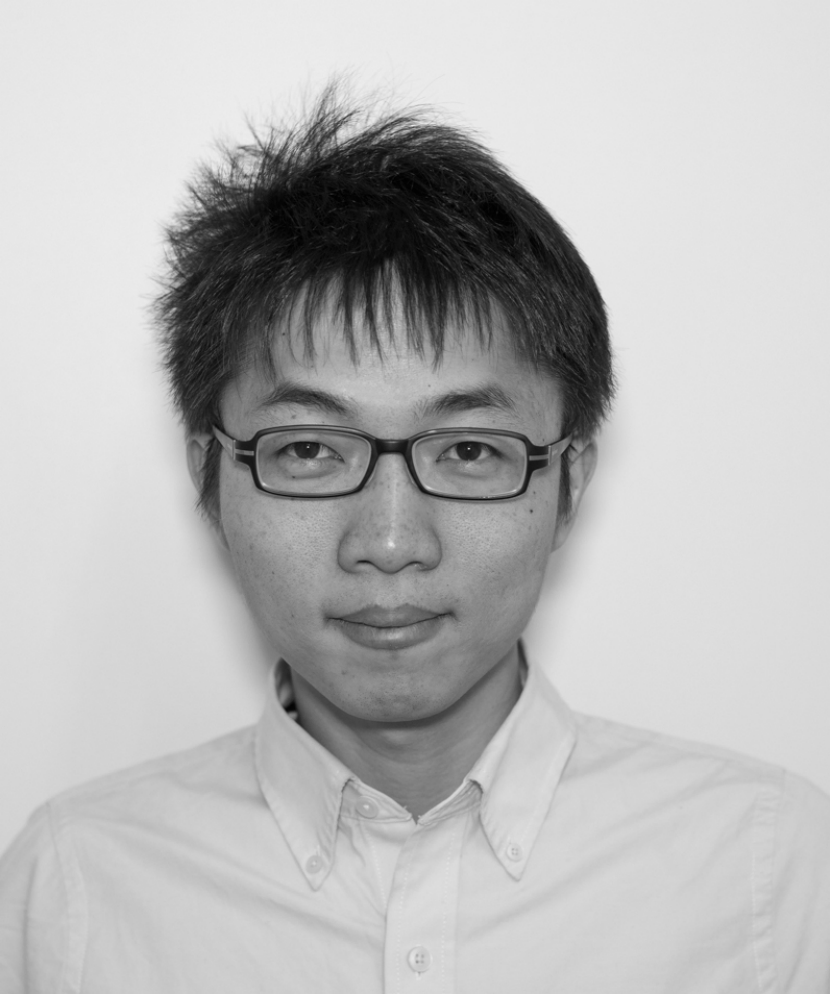}}]{Po-Sen Huang} is a research scientist at Clarifai. He received his B.S. in Electrical Engineering from National Taiwan University in 2008, and his M.S. and Ph.D. degrees in Electrical and Computer Engineering from the University of Illinois at Urbana-Champaign in 2010 and 2015, respectively. His research interests include machine learning for audio and natural language processing, with a focus on deep learning and large-scale kernel machines. He is the recipient of the Yi-Min Wang and Pi-Yu Chung Endowed Research Award from UIUC in 2014, the Starkey Signal Processing Research Student Grant, and the IBM Research Spoken Language Processing Student Grant in the International Conference on Acoustics, Speech and Signal Processing (ICASSP) 2014.
\end{IEEEbiography}
\vspace{-7mm}
\begin{IEEEbiography}[{\includegraphics[width=1in,height=1.25in,clip,keepaspectratio]{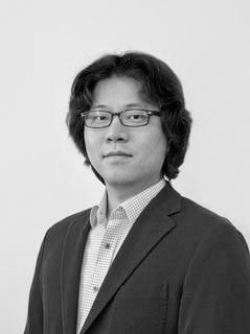}}]{Minje Kim} is a PhD candidate in the Department of Computer Science at the University of Illinois at Urbana-Champaign. Before joining UIUC, he worked as a researcher in ETRI, a national lab in Korea, from 2006 to 2011. He did his Bachelor's and Master's studies in the Division of Information and Computer Engineering at Ajou University and in the Department of Computer Science and Engineering at POSTECH in 2004 and 2006, respectively. His research focuses on machine learning algorithms applied to audio processing, stressing out the computational efficiency in the resource-constrained environments or in the applications involving large unorganized datasets. He received Richard T. Cheng Endowed Fellowship from UIUC in 2011. Google and Starkey grants also honored his ICASSP papers as the outstanding student papers in 2013 and 2014, respectively.
\end{IEEEbiography}
\vspace{-7mm}
\begin{IEEEbiography}[{\includegraphics[width=1in,height=1.25in,clip,keepaspectratio]{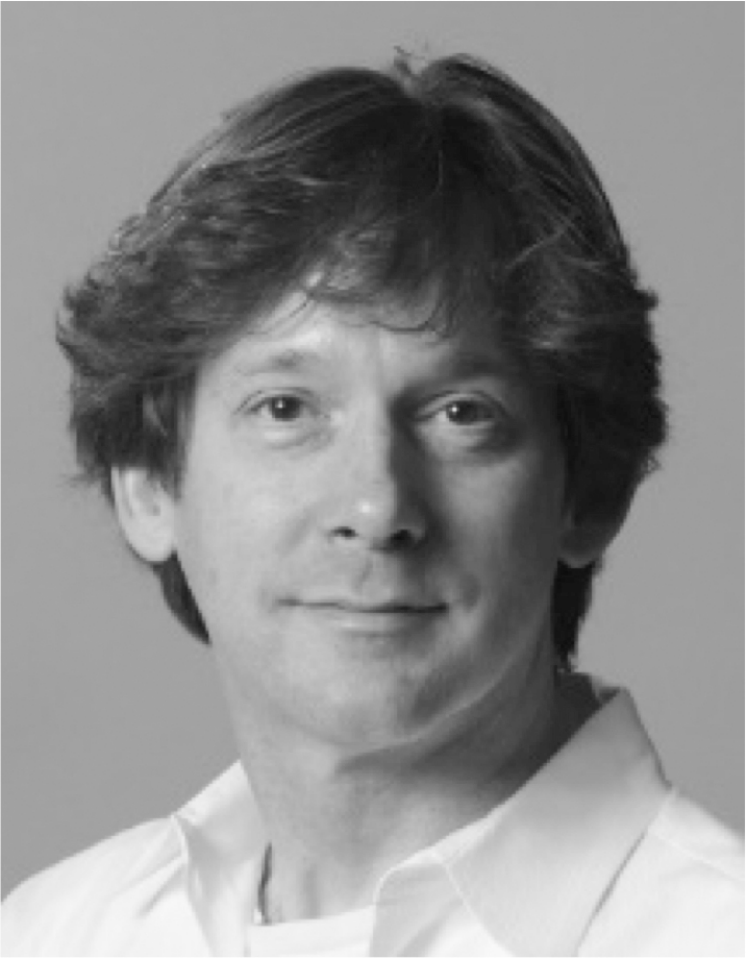}}]{Mark Hasegawa-Johnson} (M'88--SM'05) received the M.S. and Ph.D. degrees from MIT in 1989 and 1996, respectively. He is Professor of ECE at the University of Illinois at Urbana-Champaign, Full-Time Faculty in the Beckman Institute for Advanced Science and Technology, and Affiliate Professor in the Departments of Speech and Hearing Science, Computer Science, and Linguistics. He is currently a member of the IEEE SLTC, an Associate Editor of JASA, Treasurer of ISCA, and Secretary of SProSIG. He is author or co-author of 53 journal articles, 157 conference papers, 47 printed abstracts, and 5 US patents. His primary research areas are in the application of phonological concepts to audio and audiovisual speech recognition and synthesis (Landmark-Based Speech Recognition), in the application of semi-supervised and interactive machine learning methods to multimedia browsing and search (Multimedia Analytics).
\end{IEEEbiography}
\vspace{-7mm}
\begin{IEEEbiography}[{\includegraphics[width=1in,height=1.25in,clip,keepaspectratio]{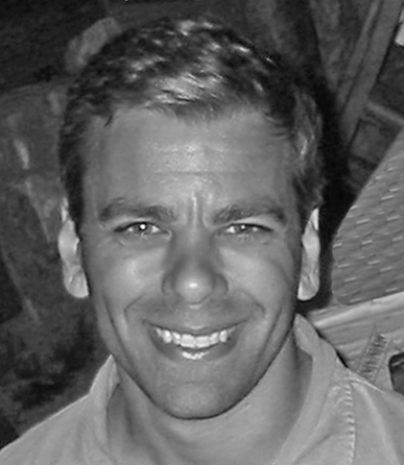}}]{Paris Smaragdis} is faculty in the Computer Science and the Electrical and Computer Science departments at the University of Illinois at Urbana-Champaign. He is also a senior research scientist at Adobe. He completed his graduate and post-doctoral studies at MIT, where he conducted research on computational perception and audio processing. Prior to the University of Illinois he was a senior research scientist at Adobe Systems and a research scientist at Mitsubishi Electric Research Labs, during which time he was selected by the MIT Technology Review as one of the top 35 young innovators of 2006. He is a member of the IEEE AASP TC and has previously been the chair of the IEEE MLSP TC, and is an IEEE Fellow. Paris' research interests lie in the intersection of machine learning and signal processing, especially as they apply to audio problems.
\end{IEEEbiography}

\end{document}